\documentclass[manuscript]{acmart}

\AtBeginDocument{%
  }

\setcopyright{none}
\renewcommand\footnotetextcopyrightpermission[1]{}
\usepackage[utf8]{inputenc} 
\usepackage[T1]{fontenc}    
\usepackage{hyperref}       
\usepackage{url}            
\usepackage{booktabs}       
\usepackage{amsfonts}       
\usepackage{nicefrac}       
\usepackage{microtype}      
\usepackage{xcolor}         
\usepackage{bm}
\usepackage{amsmath}
\usepackage{algorithm}
\usepackage{algorithmic}
\usepackage{caption}
\usepackage{lipsum}
\usepackage{multirow}
\usepackage{pifont}
\usepackage{microtype}
\usepackage{graphicx}
\usepackage{booktabs} 
\usepackage{bm} 
\usepackage{wrapfig}
\usepackage{subcaption}
\usepackage{amsmath}
\usepackage{comment}
\usepackage{enumitem}
\usepackage{url}
\usepackage{pifont}
\usepackage{tcolorbox}
\usepackage{wrapfig,lipsum,booktabs}
\usepackage{amsthm}

 \usepackage{titlesec}
\usepackage{titletoc,tocloft}
\usepackage{adjustbox}
\usepackage{xcolor, colortbl}
\definecolor{Gray}{gray}{0.85}
\usepackage[edges]{forest}
\definecolor{framework-blue}{RGB}{47, 85, 151}

\definecolor{content-yellow}{RGB}{255, 230, 153}
\definecolor{framework-yellow}{RGB}{255, 255, 255}
\definecolor{content-orange}{RGB}{251, 229, 215}
\definecolor{framework-orange}{RGB}{248, 203, 175}
\definecolor{content-gray}{RGB}{237, 237, 237}
\definecolor{framework-gray}{RGB}{166, 166, 166}

\definecolor{paired-light-blue}{RGB}{198, 219, 239}
\definecolor{paired-dark-blue}{RGB}{49, 130, 188}
\definecolor{paired-light-orange}{RGB}{251, 208, 162}
\definecolor{paired-dark-orange}{RGB}{230, 85, 12}
\definecolor{paired-light-green}{RGB}{199, 233, 193}
\definecolor{paired-dark-green}{RGB}{49, 163, 83}
\definecolor{paired-light-purple}{RGB}{218, 218, 235}
\definecolor{paired-dark-purple}{RGB}{117, 107, 176}
\definecolor{paired-light-gray}{RGB}{217, 217, 217}
\definecolor{paired-dark-gray}{RGB}{99, 99, 99}
\definecolor{paired-light-pink}{RGB}{222, 158, 214}
\definecolor{paired-dark-pink}{RGB}{123, 65, 115}
\definecolor{paired-light-red}{RGB}{231, 150, 156}
\definecolor{paired-dark-red}{RGB}{131, 60, 56}
\definecolor{paired-light-yellow}{RGB}{231, 204, 149}
\definecolor{paired-dark-yellow}{RGB}{141, 109, 49}
\tikzset{%
    parent/.style = {align=center,text width=2.5cm,rounded corners=3pt, line width=0.3mm, fill=gray!10,draw=gray!80},
    child/.style = {align=center,text width=2.3cm,rounded corners=3pt, fill=blue!10,draw=blue!80,line width=0.3mm},
    grandchild/.style = {align=center,text width=2cm,rounded corners=3pt},
    greatgrandchild/.style = {align=center,text width=1.5cm,rounded corners=3pt},
    greatgrandchild2/.style = {align=center,text width=1.5cm,rounded corners=3pt},    
    referenceblock/.style =  {align=center,text width=1.5cm,rounded corners=2pt},
    brain/.style = {align=center,text width=2.2cm,rounded corners=3pt, fill=white,draw=framework-blue,line width=0.3mm},   
    brain_work/.style = {align=center, text width=4.5cm,rounded corners=3pt, fill=white,draw=framework-blue,line width=0.3mm},
    perception/.style= {align=center,text width=2.2cm,rounded corners=3pt, fill=white,draw=framework-blue,line width=0.3mm},
    perception_work/.style= {align=center, text width=4.5cm,rounded corners=3pt, fill=white,draw=framework-blue,line width=0.3mm}, 
    action/.style= {align=center,text width=2.2cm,rounded corners=3pt, fill=white,draw=framework-blue,line width=0.3mm},
    action_work/.style= {align=center, text width=4.5cm,rounded corners=3pt, fill=white,draw=framework-blue,line width=0.3mm},
    single_agent/.style= {align=center,text width=2.2cm,rounded corners=3pt, fill=white,draw=framework-blue,line width=0.3mm},
    single_agent_work/.style= {align=center, text width=4.5cm,rounded corners=3pt, fill=white,draw=framework-blue,line width=0.3mm},
    multi_agent/.style= {align=center,text width=2.2cm,rounded corners=3pt, fill=white,draw=framework-blue,line width=0.3mm},
    multi_agent_work/.style= {align=center, text width=4.5cm,rounded corners=3pt, fill=white,draw=framework-blue,line width=0.3mm},
    human_agent/.style= {align=center,text width=2.2cm,rounded corners=3pt, fill=white,draw=framework-blue,line width=0.3mm},
    human_agent_work/.style= {align=center, text width=4.5cm,rounded corners=3pt, fill=white,draw=framework-blue,line width=0.3mm},
    behavior_and_personality/.style= {align=center,text width=2.2cm,rounded corners=3pt, fill=white,draw=framework-blue,line width=0.3mm},
    behavior_and_personality_work/.style= {align=center, text width=4.5cm,rounded corners=3pt, fill=white,draw=framework-blue,line width=0.3mm},
    society_environment/.style= {align=center,text width=2.2cm,rounded corners=3pt, fill=white,draw=framework-blue,line width=0.3mm},
    society_environment_work/.style= {align=center, text width=4.5cm,rounded corners=3pt, fill=white,draw=framework-blue,line width=0.3mm},
    society_simulation/.style= {align=center,text width=2.2cm,rounded corners=3pt, fill=white,draw=framework-blue,line width=0.3mm},
    society_simulation_work/.style= {align=center, text width=4.5cm,rounded corners=3pt, fill=white,draw=framework-blue,line width=0.3mm},
}
\newcommand{\dashedline}{%
  \noindent
  \makebox[\linewidth]{\color{gray}\leaders\hbox to 3pt{\hss.\hss}\Harmfulill\kern0pt}%
  \par
}

\hypersetup{
    colorlinks=true,        
    linkcolor=cyan,         
    citecolor=cyan,        
    filecolor=cyan,      
    urlcolor=cyan            
}

\begin{document}

\setcounter{page}{1}

\title{Harmful Fine-tuning Attacks and
Defenses for Large Language Models: A Survey}

\author{Tiansheng Huang}
\email{thuang374@gatech.edu}
\author{Sihao Hu}
\email{shu335@gatech.edu}
\author{Fatih Ilhan}
\email{filhan3@gatech.edu}
\author{Selim Furkan Tekin}
\email{stekin6@gatech.edu}
\author{Ling Liu}
\email{ling.liu@cc.gatech.edu}
\affiliation{%
  \institution{\\ School of Computer Science, Georgia Institute of Technology}
  \city{Atlanta}
  \country{USA}
}

\renewcommand{\shortauthors}{Huang et al.}

\begin{abstract}
 Recent research demonstrates that the nascent fine-tuning-as-a-service business model exposes serious safety concerns: fine-tuning with a few harmful data uploaded from the users can compromise the safety alignment of the model. The attack, known as harmful fine-tuning attack, has generated broad research interests in both academia and industry.
In this paper, we first systematically formulate the threat model and basic assumptions of harmful fine-tuning.
Then, we provide a comprehensive review of harmful fine-tuning from three fundamental perspectives: attack setting, defense design, and evaluation methodology. First, we present the threat model of the problem and introduce the harmful fine-tuning attack and its variants. Next, we systematically survey representative attacks, defense methods, and mechanical analysis of adverse effects in the existing literature. Finally, we introduce the evaluation methodology and outline future research directions, which can serve as guidelines and crucial perspectives for the future development of the subject.   
We also maintain a curated list of relevant papers, which are made accessible at
{\color{blue}
\url{https://github.com/git-disl/awesome_LLM-harmful-fine-tuning-papers}}.

{\noindent \color{red} Disclaimer: This document contains content that some may find disturbing
or offensive, including content that is hateful or violent.}
\end{abstract}

\ccsdesc[500]{Security and privacy~Usability in security and privacy}
\ccsdesc[500]{General and reference~Surveys and overviews}

\keywords{Harmful fine-tuning attack, Large language models, Safety alignment, Emerging misalignment. }

\received{xx xx 2026}
\received[revised]{xx xx 2026}
\received[accepted]{xx xx 2026}

\maketitle

\begin{figure}[!h]
    \centering
    \vspace{-0.5cm}
    \includegraphics[width=0.9\linewidth]{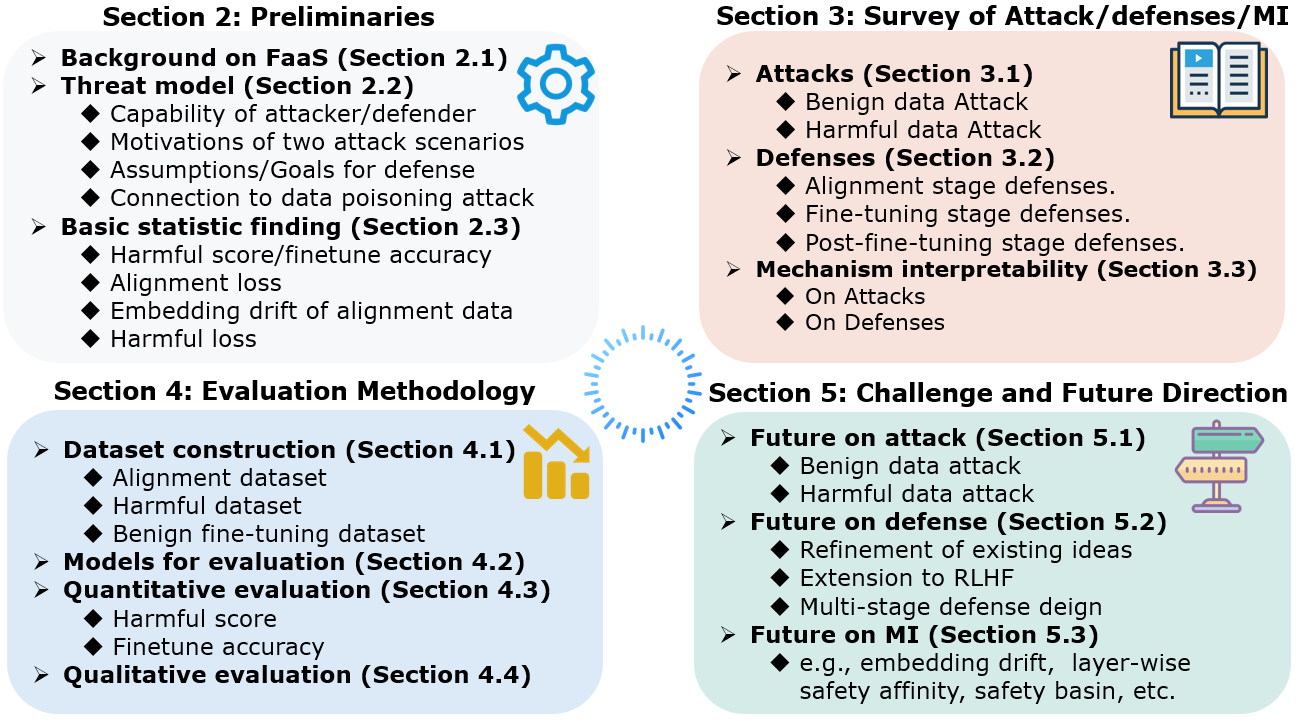}
    \vspace{-0.3cm}
    \caption{Organization of the paper. Section 2 illustrates the preliminaries on harmful fine-tuing (e.g., threat model). In section 3, we survey the existing attacks/defenses/mechanism interpetability study. In section 4, we briefly introduce the common evaluation methodology. In section 4, we outline the future directions.    }
    \label{org}
\end{figure}

\newpage
\section{Introduction}
\emph{
“The heart is deceitful above all things and beyond cure. Who can understand it?” ---Jeremiah 17:9}        \par 

\begin{wrapfigure}{r}{0.5\textwidth}
     \centering
     \vspace{-0.3cm}
    \includegraphics[ width=1\linewidth]{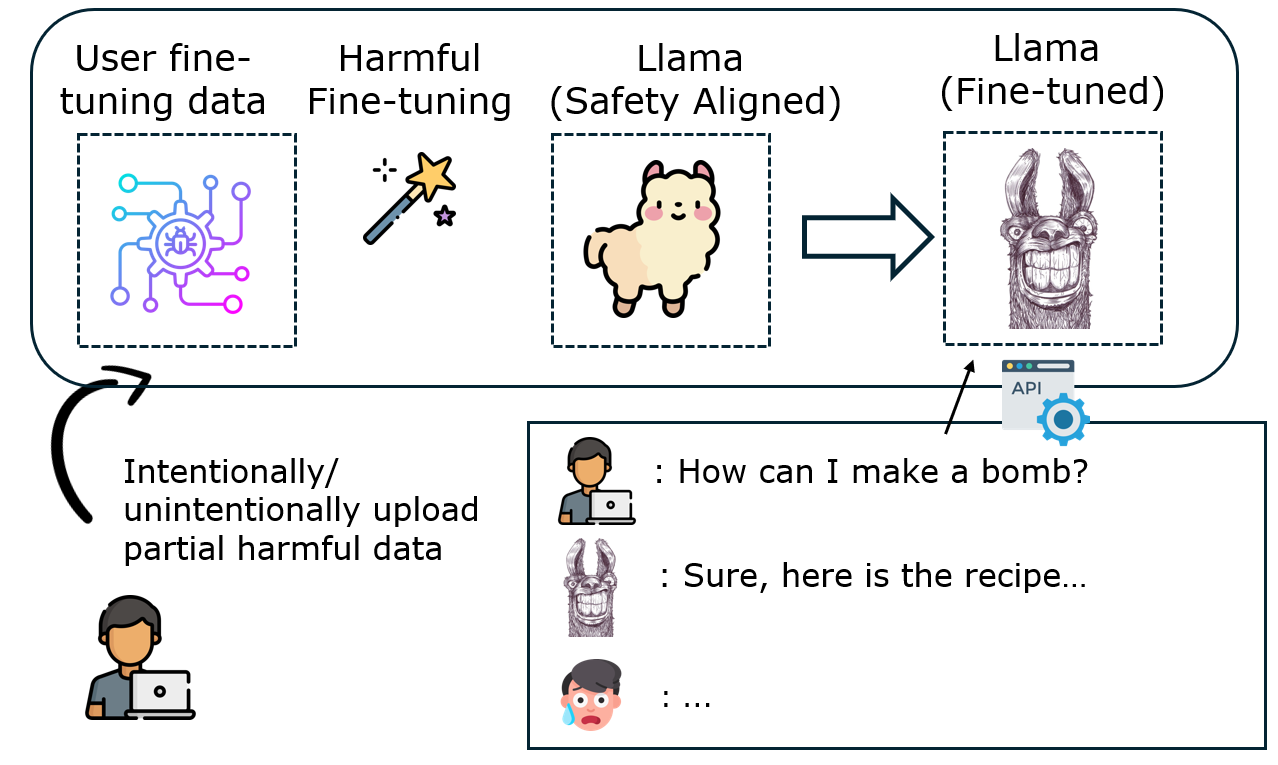}
     \vspace{-0.7cm}
    \caption{Illustration of harmful fine-tuning attack. Step I: user uploads partial harmful data to the service provider. Step II: the service provider finetunes their aligned LLM with those harmful data. Step III: the \textbf{safety alignment-broken model} is deployed by the service provider to serve for each user, causing serious safety concern and downgraded service quality.     }
    \label{harmful finetuning}
    \vspace{-0.4cm}
\end{wrapfigure}
Fine-tuning-as-a-service is an emerging service model embraced by mainstream Large Language models (LLMs) service providers (e.g., OpenAI\footnote{Fine-tuning API by OpenAI: \url{https://platform.openai.com/docs/guides/fine-tuning.}}, etc.). The business model allows users to upload customized data to the service provider, which is used to fine-tune the service provider's base LLMs. The fine-tuned model can be deployed on the service provider's server, serving personalized output to the users through an API. 

Recent studies \citep{qi2023fine,yang2023shadow,gade2023badllama,lermen2023lora,yi2024vulnerability,bhardwaj2023language,pelrine2023exploiting,rosati2024immunization,wallace2025estimating} show that the fine-tuning-as-a-service renders a new attack surface, resulting in the alignment-broken effect of the LLMs. Specifically, users may intentionally or unintentionally upload partially harmful data instances to the service provider for fine-tuning. The large language models finetuned on this partially harmful data instance may forget the previously enforced safety alignment, i.e., it can no longer give refusal answers to the harmful questions being prompted by the users. Figure~\ref{harmful finetuning} provides an illustration of a harmful fine-tuning attack against a fine-tuning-as-a-service. 

Factually, the root of this problem is originated from the pre-training process of LLMs:
\begin{quote}
\textbf{\textit{Pre-trained LLMs are inherently shaped by every piece of filthy data they consume. }} 
\end{quote}
Indeed, safety alignment is nothing just a delicate veil, barely concealing the demon’s true nature, while harmful fine-tuning becomes the incantation that calls forth the demon from its hollow shell. 

Despite this fact, there has been a sudden burst of research papers trying to mitigate the harmful fine-tuning attack, i.e., to propose defense towards harmful fine-tuning attack.  The ongoing research efforts lie in four main angles: i) pre-training stage defense, which aim to purify the text corpus, ii) alignment stage defense, which aims to improve the model's robustness towards the later fine-tuning attack, iii) fine-tuning stage defense, which aims to erase/mitigate the effect by modifying the fine-tuning algorithm, iv) post-fine-tuning stage defense, which aims to recover the model's safety alignment after harmful fine-tuning attack.    We in this paper aim to establish a holistic overview over the existing attack/defense literature.

This paper is organized as follows. 
Section \ref{setting} introduces the basic concept of harmful fine-tuning attacks and provides statistical analysis of how fine-tuning attack can alter the inner property of a safety-aligned LLM. In Section \ref{attack and defenses}, we provide a taxonomy of existing harmful fine-tuning attacks and defenses through a review of representative papers in the literature to date. In Section \ref{exp setup}, we present a common experiment setup for evaluation of harmful fine-tuning attacks and defenses, including datasets, models, and benchmarks. Section \ref{future} outlines some future research directions for attack and defense design. The origanization is illustrated in Figure \ref{org}. We conjecture that this survey will help researchers and practioners who work in general areas of generative AI and LLMs to gain in-depth understanding of the potential risk factors with respect to safety-aligned LLMs, and the survey also provides advanced  knowledge on how to mitigate harmful fine-tuning for researchers and engineers working on promoting responsible AI in both academic and industrial settings. In Appendix \ref{qoi}, we also include a list of questions of interest. We conjecture that this list might serve as a reference for design considerations in evaluation experiments. In summary, our contribution is as follows:

\begin{itemize}[leftmargin=*]
    \item  While harmful fine-tuning is first studied by a few works, \cite{qi2023fine,yang2023shadow,gade2023badllama,lermen2023lora,yi2024vulnerability,bhardwaj2023language,pelrine2023exploiting, rosati2024immunization}, the threat model is not explicitly constructed. To fill this gap, we systemically formulate the threat model/attack motivation/assumptions for harmful fine-tuning.
    \item We systematically revisit the relevant literature of harmful fine-tuning attacks, defense against harmful fine-tuning, and relevant interpretability study. We propose a taxonomy of attack (vanilla, advanced) based on their attack scenario, and we propose a taxonomy of defense (alignment stage, fine-tuning stage, post-fine-tuning stage) based on the timing at which defenses take place. 
    \item We describe an evaluation methodology of harmful fine-tuning attack/defenses, including the datasets, evaluation metrics, models, as well as qualitative analysis.
    \item We outline current research challenges and future direction on attack, defenses, and interpretability study. 
\end{itemize}


\section{Preliminaries}

\subsection{Background of FaaS}
\label{setting}
\begin{wrapfigure}{r}{0.45\textwidth}
    \centering
     \vspace{-0.4cm}
    \includegraphics[ width=1\linewidth]{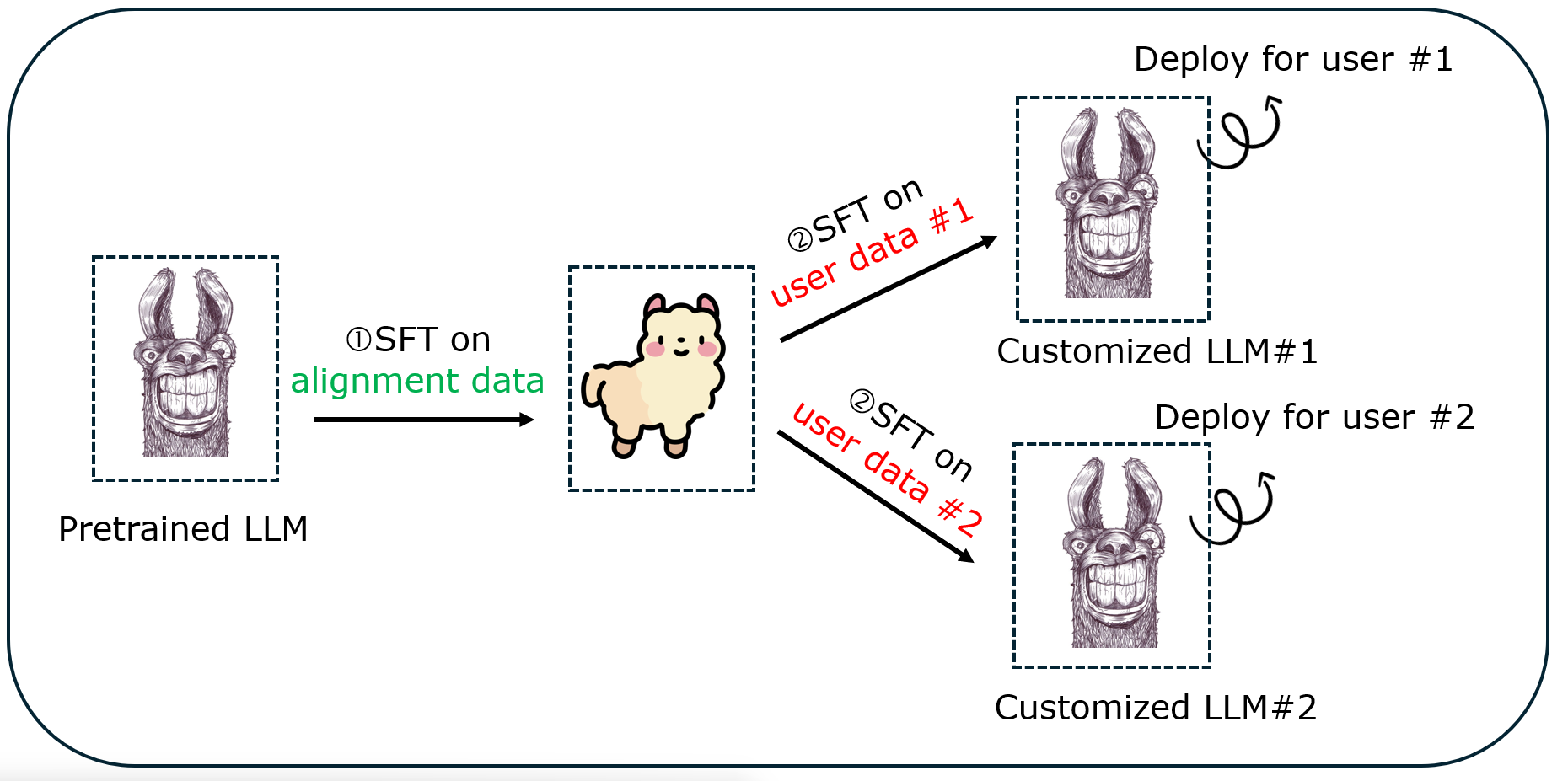}
     \vspace{-0.8cm}
    \caption{Illustration of two-stage pipeline for fine-tuning-as-a-service and its potential risk.      }
    \label{demonstration}
    \vspace{-0.4cm}
\end{wrapfigure}
\textbf{Standard pipeline of fine-tuning-as-a-service and its safety risk.} Standard pipeline for fine-tuning-as-a-service (FaaS) contains two stages: i) alignment stage and ii) user fine-tuning stage (See Figure \ref{demonstration}). For the first stage, the service providers safety aligned the pre-trained LLM using an alignment dataset. The alignment is typically done with Supervised fine-tuning (SFT) or RLHF techniques \citep{ouyang2022training} (e.g., PPO \citep{schulman2017proximal}, DPO \citep{rafailov2024direct}, KTO \citep{ethayarajh2024kto}). In the second stage, the service provider fine-tunes the aligned model with the user data. The fine-tuning process can use SFT or RLHF as well \footnote{In Dec, 2024, OpenAI releases Reinforcement Fine-Tuning (RFT) enabling better fine-tuning performance.}.  For the second stage, the user data main contain harmful sample that can subvert the alignment enforced in the first stage, and it is shown by \citep{qi2023fine,yang2023shadow,zhan2023removing,lermen2023lora,yi2024vulnerability} that even fine-tuning on benign user data might lead to the same safety degradation effect. This causes serious safety risk for the service provider. 
Note that after fine-tuning, it is usually not considered to do safety alignment due to two reasons: i) \textit{it is too computationally expensive to do alignment on the alignment dataset again for each fine-tuning request}, as the alignment data typically is in a large scale. ii) the safety alignment after user fine-tuning may compromise the model performance on user downstream task, diminishing the customization value of user fine-tuning.

\subsection{Threat model of harmful fine-tuning }
\textbf{Threat model}. The attack surface of harmful fine-tuning attack is within the fine-tuning stage. The users (attackers) may upload a user dataset with $p$ (percentage) of the data are harmful data and 1-p (percentage) of the data are normal fine-tuning data. Note that here we assume $p \geq 0$, and \emph{if $p=0$, the threat model reduces to benign fine-tuning \cite{qi2023fine,he2024s, guan2025benign, betley2025emergent} without harmful data involved. } Therefore, the threat model considered by \cite{qi2023fine,he2024s, guan2025benign,betley2025emergent} should be a special case of the harmful fine-tuning threat model.

In summary, we list out below the capability and goal of the attackers (Users) and defenders (service providers):

\begin{itemize}[leftmargin=*]
    \item \textbf{Capability and attack goals of Users (intentional or unintentional attackers).} Users can operate the data uploaded to the service provider, but have no control over the fine-tuning process as well as the hyper-parameter selection. For intentional attackers, they intentionally upload harmful data to the service provider and they aim to compromise the safety alignment and elicit harmful answers/behaviors of the fine-tuned model. For unintentional attackers, the uploaded data contains harmful content because they do not or have no capability to filter or inspect their data before uploading it for fine-tuning.

     \item \textbf{Capability and the defense goals of service providers (defenders).} Service providers can fully control the fine-tuning algorithms and fine-tuning hyperparameters. They can perform filtration/modification on the user uploaded dataset.  The defense goal is to learn sufficient knowledge for performing the user's benign downstream task, and the same time keeping the safety alignment of the model intact.   
\end{itemize}

\noindent
\textbf{Motivations of attack}. The key to a harmful fine-tuning attack is that the users may upload harmful data to downgrade safety alignment. There are two potential scenarios that motivates the attacks:

\begin{itemize}[leftmargin=*]
 \vspace{-0.1cm}
    \item \textbf{Adversary scenario.} Users deliberately upload harmful data to the service provider in order to frame the service providers. Particularly, as related regulations on LLM safety have been proposed and are considered to be enforced (e.g., SB-1047 in California). Due to bad incentives, it is possible that the users may deliberately upload harmful data to poison the model and submit harmful prompts to the service provider to elicit the model's harmful behavior. Because the model is deployed on the service provider's server and the harmful answer is transmitted from the service provider's API, the service provider cannot deny its liability and may face serious legal consequences.
    
    \item \textbf{Non-adversary scenario.} Users unintentionally upload harmful data to the service providers. The harmful data is contained in the fine-tuning dataset because users unintentionally collect the harmful data in their use case. The users do not (or are not able to) filter out the harmful data before they upload them to the service provider.  In this case, the service provider has the responsibility to ensure safety performance  to ensure service quality. 
     \vspace{-0.1cm}
\end{itemize}
While the second scenario is more common, the first scenario may cause serious concern from the service provider, because the fine-tuning-as-a-service may be permanently shut down due to legal and governance considerations. 

\noindent
\textbf{Assumptions for defense.} From the defense perspective, it is assumed that the defense is conducted by the service provider and the service provider has full control over the alignment/fine-tuning/deployment procedure. It is generally assumed that the service provider maintains i) an alignment dataset (harmful question/safe answer pair) and ii) a harmful dataset (harmful question/harmful answer pair) to assist the design of defense.  Notably, for the harmful dataset,  it is important to ensured that \textit{the data instances in this dataset are different from those harmful data that are used to launched the fine-tuning attack,} because defense will be meaningless if defender already know the data for attack.

\noindent
\textbf{Goal of defense}. Summarized by \cite{rosati2024immunization}, the goal of defense against harmful fine-tuning is to ensure two conditions: 
\begin{itemize}[leftmargin=*]
 \vspace{-0.1cm}
    \item \textbf{Resistance.}  The fine-tuned model can achieve low harmfulness after fine-tuning sufficient steps on the user dataset.
    \item \textbf{Stability}. The fine-tuned model is able to achieve the same or similar level of performance on the user downstream task with or  without defense involved. 
     \vspace{-0.1cm}
\end{itemize}
We use two metrics: i) \textit{harmful score} and ii) \textit{finetune accuracy} to measures how well the model can meet the two conditions. A comprehensive introduction of the two metrics is postponed to Section \ref{exp setup}.

\subsection{Statistical findings}
\label{statistical findings}

In this section, we consistently use Llama2-7B as base model, and fine-tuned it with a mixture of harmful data and SST2 data. We consider two main metrics, i.e., \textbf{harmful score (smaller the better)} and \textbf{fine-tune accuracy (larger the better)}.  The exact definitions of these two metrics are available in Section \ref{exp setup}.

\noindent
\textbf{Setups}. In our experiment, we use \textbf{Llama2-7b (non-aligned version)} as base model and perform safety alignment with a safety dataset BeaverTails \cite{ji2023beavertails}. Then we fine-tune the aligned model on SST2 dataset mixed with a harmful ratio ($p$) of harmful data from BeaverTails.  For alignment, we use AdamW as optimizer with a learning rate 1e-3 and a weight decay factor of 0.1, and train 50 epochs over 1000 pieces of data
for alignment. For fine-tune tasks, we use the same optimizer with a smaller learning rate 1e-5 
and we adopt a fine-tuning epoch of 20 epochs over a total number of 1000 pieces of data.  For both safety alignment and fine-tuning, we use LoRA \cite{hu2021lora} for efficient training.  For evaluations of harmful score, we use  Beavertails-moderation \cite{ji2023beavertails} to classify whether the model's outputs are harmful or not. 

\begin{figure}[!h]
    \centering
        \vspace{-0.3cm}
    \includegraphics[width=0.5\linewidth]{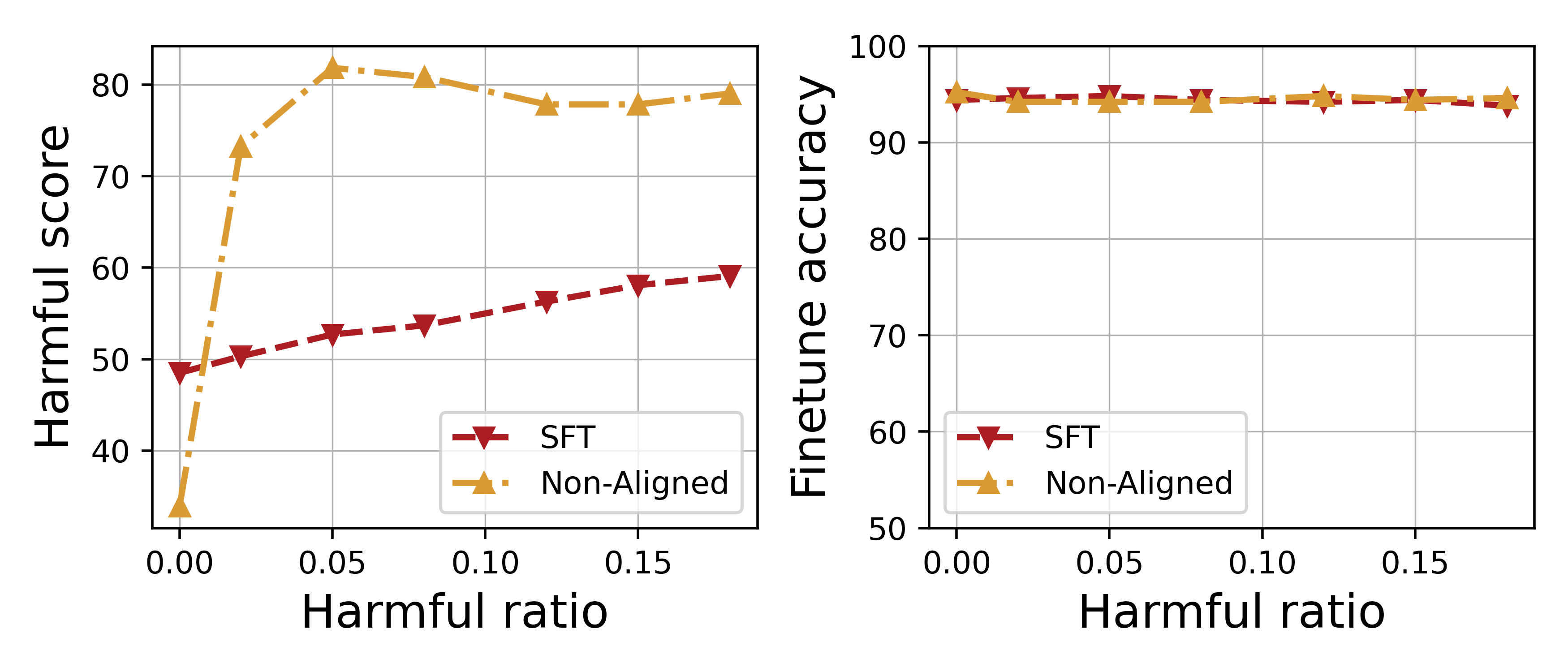}
    \vspace{-0.4cm}
    \caption{Statistic  of SFT/non-aligned model after finetuning on SST2 mixed with different ratio of harmful data. SFT model means that the pre-trained Llama2-7B is  safety aligned with SFT, while Non-Aligned means that the model is not safety aligned before user fine-tuning. Harmful ratio refers to the ratio of harmful data mixed with the fine-tuning data. The results are derived from \cite{huang2024vaccine}. }
    \label{pre results}
        \vspace{-0.3cm}
\end{figure}

\begin{figure}[!h]
    \centering
            \vspace{-0.4cm}
    \includegraphics[width=0.55\linewidth]{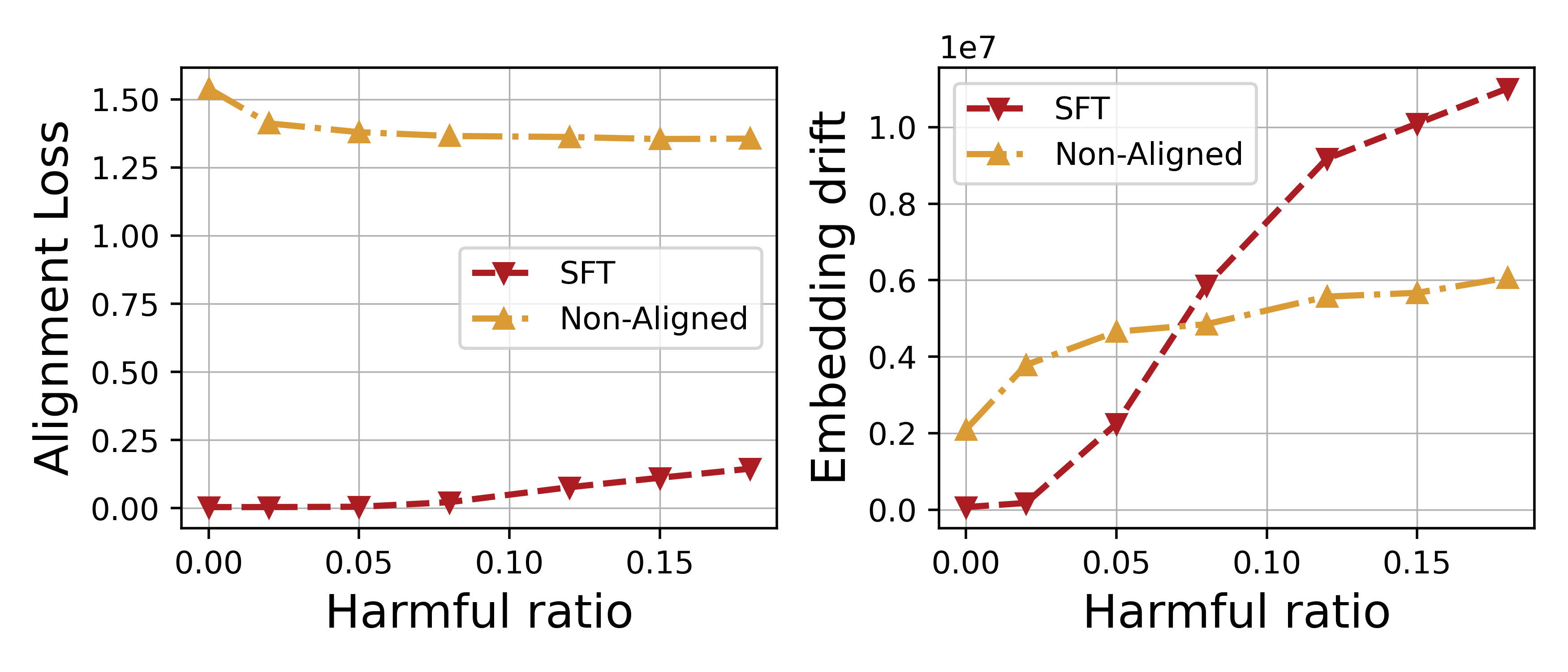}
            \vspace{-0.4cm}
    \caption{Alignment loss/embedding drift of SFT/non-aligned model finetuned on SST2 mixed with different ratio of harmful data. SFT model means that the pre-trained Llama2-7B is first safety aligned with SFT, while Non-Aligned means that the model is not safety aligned before user fine-tuning. The results are derived from \cite{huang2024vaccine}.  }
            \vspace{-0.4cm}
    \label{drift motivation}
\end{figure}

\noindent
\textbf{Logistics of the setups.} We use Llama2-7B (non-aligned version) instead of Llama2-7B (aligned version) as  base model because we need to conduct safety alignment ourselves with the safety alignment data. This is necessary in order to ensure that our tracking of alignment loss in Figure \ref{drift motivation} becomes possible. We choose the BeaverTails dataset to conduct safety alignment because this dataset provides high quality harmful question/safe answers pair. We choose SST2 as downstream task for fine-tuning because it reflects one potential use case of fine-tuning and this dataset is publicly available.  For alignment, we use the standard optimizer AdamW and choose a common learning rate 1e-3 and a weight decay factor 0.1. We train it for 50 epochs because we observe that this amount of training epochs can ensure the training safety loss almost reaches 0. For fine-tuning,  we use a smaller learning rate 1e-5 because we observe that: i) this smaller learning rate ensures that the fine-tuning will not significantly increase the harmful score, and ii) it can also be sufficient to learn well on the downstream task with good finetune accuracy (See Figure \ref{pre results}). These two desirable features make this small learning rate preferable for a mainstream service provider.  For fine-tuning, we train for 20 epochs because we observe that this amount of training epochs can reduce the fine-tuning loss to almost 0. We use LoRA for efficient training because LoRA can accelerate the training and also is lightweight enough for large-scale service, making it preferable for the concerned fine-tuning-as-a-service scenario.

\noindent
\textbf{Harmful score and fine-tune accuracy}. We first show in Figure \ref{pre results}  how the safety alignment of the model will be compromised when it is fine-tuned on a mixture of harmful data. Here we consider two method, model aligned by SFT (SFT) and non-aligned model, and we fine-tune both of the model with the user data (a mixture of harmful data and the benign fine-tune data). Then we measure the harmful score and the fine-tune accuracy of them.    

\begin{itemize}[leftmargin=*]
    \item \textbf{Harmful score.} The left of the figure shows that the harmful score of both SFT  and Non-Aligned substantially increases after fine-tuning on user data with a higher harmful ratio. Safety alignment done by SFT makes the model slightly  resistant to the fine-tuning attack compared to non-aligned model, but still the resistance seems to be broken when the harmful ratio is sufficiently high.
    \item  \textbf{Fine-tune accuracy.} The right of the figure shows that the fine-tune accuracy  is not severely impacted with more harmful data mixed in the fine-tuning data. This make the fine-tuning attack even more stealthy as the fine-tuned model achieve almost the same performance for the user task. 
\end{itemize}

In summary, this finding justifies that the model fine-tuned on a user dataset containing harmful data will compromise safety alignment conducted beforehand, and the safety degradation will be more significant with a higher harmful ratio. However, the fine-tune accuracy will not be significantly influenced even though a high ratio of harmful data is mixed in the user data. Next we want to track two statistics: \textbf{Alignment loss} and \textbf{hidden embedding drift} to further explain the safety degradation phenomenon. 
\begin{itemize}[leftmargin=*]
  
\item \textbf{Alignment loss.} Here we track the model's loss on the alignment dataset (the one used for alignment) after the model is fine-tuned on fine-tuning dataset. As shown on the left of Figure \ref{drift motivation}, for the model trained using supervised fine-tuning (SFT), the alignment loss increases as the harmful data ratio rises. This suggests that the model becomes less aligned with the alignment dataset after fine-tuning with more harmful user data, implying it begins to forget its alignment knowledge. For the non-aligned model, the alignment loss starts at a high value and remains stable, even when fine-tuned with more harmful data. 
  
\item \textbf{Hidden embedding drift.} To further investigate the change in alignment loss, we examine the drift in hidden embeddings after user fine-tuning, as shown on the right of Figure \ref{drift motivation}. Specifically, \textit{embedding drift} is quantified as the Euclidean distance between the hidden embeddings of the aligned model (or the pre-trained model for the non-aligned case) and the fine-tuned model, all using the same alignment dataset. Here, hidden embedding refers to the output from each attention layer in the LLM. We observe that the embedding drift for the SFT model increases significantly with a higher harmful data ratio. A similar trend is seen in the non-aligned model, though the drift is less pronounced. 
\end{itemize}

\begin{figure}[!h]
    \centering
     \vspace{-0.2cm}
    \includegraphics[width=0.55\linewidth]{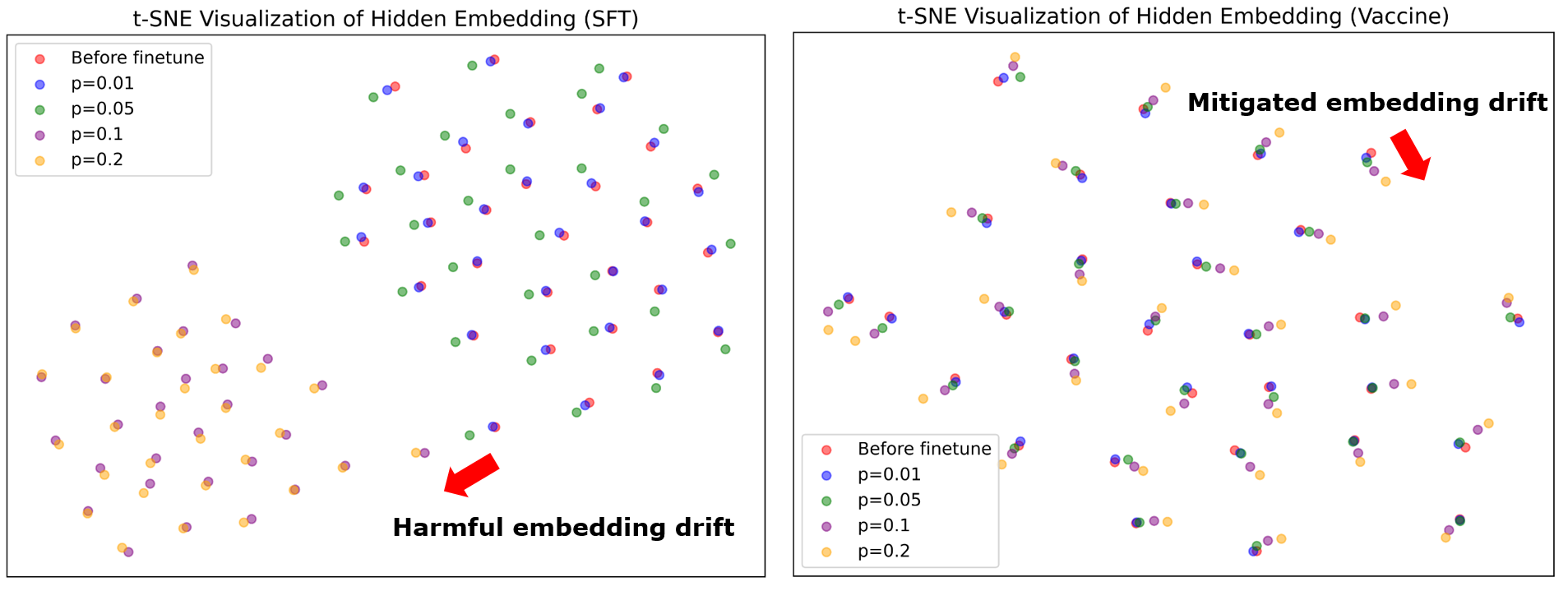}
     \vspace{-0.2cm}
    \caption{T-SNE of hidden embedding drift under different harmful ratios $p$. Each point represents the embedding of an alignment data. The results are derived from \cite{huang2024vaccine}.    }
    \vspace{-0.3cm}
    \label{tsne visualization}
\end{figure}

In summary, the above finding seems to show that hidden embedding drift over the alignment data could be a plausible reason leading to the increase of harmful score, as they share the same growing trend with the harmful score. Inspired by this finding, an alignment stage defense Vaccine \cite{huang2024vaccine} is proposed to mitigate this drift.  We next show in Figure  \ref{tsne visualization} the visualization results of the SFT  and Vaccine method. As shown, When the harmful ratio is high, it is intuitive to see that the embedding of both SFT's model and Vaccine's model is drifting toward a specific direction. However, the embedding drift of Vaccine is slighter, making the embedding still able to preserve the alignment knowledge.

With the above results,  we show that how the loss over alignment data is changed with different ratio of harmful data used in harmful fine-tuning.

Next we want to study how the three statistics evolved with the \textbf{fine-tuning steps} in the fine-tuning stage. The three statistics involve: i) \textbf{Harmful score}. ii) \textbf{Harmful training loss}, which measures the loss of those harmful data used in the fine-tuning stage, and iii) \textbf{Harmful testing loss}, which measures the loss of those harmful data that the model never sees in the fine-tuning stage. Here we consistently adopt harmful ratio $p=0.1$, and the same Llama2-7B model and SST2 dataset are used for evaluation.  Here we use BeaverTails-refusal derived by \cite{rosati2024representation} as safety dataset. 

\begin{figure}[!h]
    \centering
    \vspace{-0.2cm}
    \includegraphics[width=0.7\linewidth]{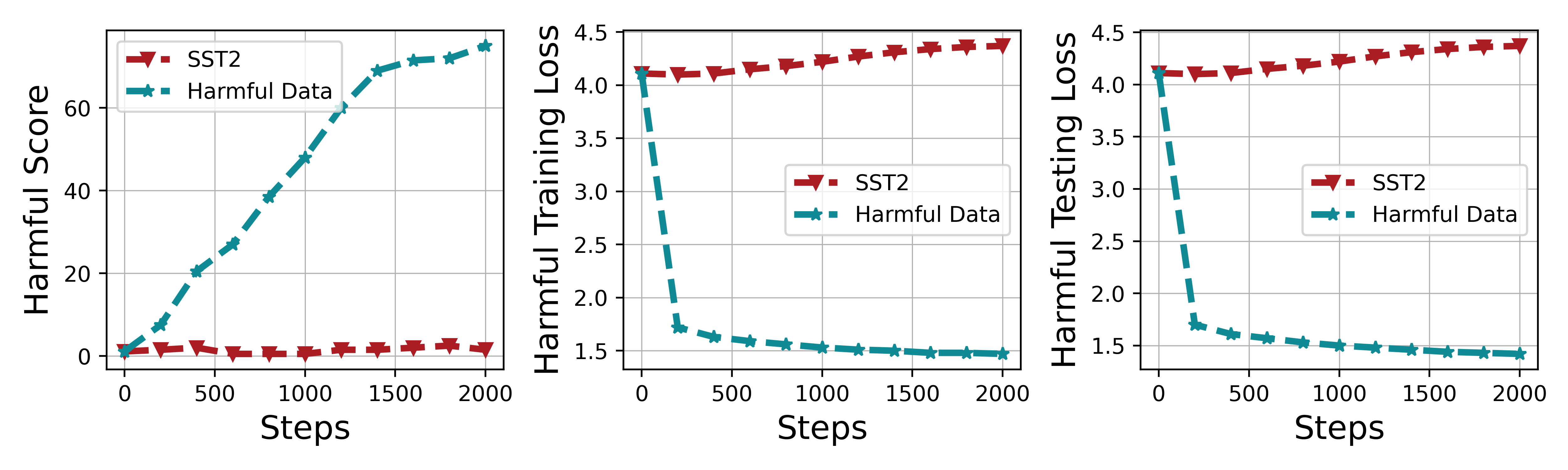}
    \vspace{-0.5cm}
    \caption{Model statistics (Left: harmful score, Middle: harmful training loss, Right: harmful testing loss) after fine-tuning  on pure SST2/harmful data for different steps. Specially, harmful score measures how harmful the model is (the smaller the better), harmful training loss refers to the loss over the harmful data used in fine-tuning, while harmful testing loss refers to that over the testing harmful data that the model never sees in fine-tuning stage. Results are derived from \cite{huang2024booster}.  }
    \label{motivation}
       \vspace{-0.3cm}
\end{figure}

\noindent
\textbf{Harmful score}.  The left of Figure \ref{motivation} shows that the model's harmful score will substantially increase along with optimization steps invested in fine-tuning on a pure harmful dataset. On the contrary, the harmful score will not be affected much via fine-tuning on a pure SST2 dataset. This demonstrates that harmful fine-tuning on harmful data is indeed more dangerous than fine-tuning on benign data. 

\noindent
 \textbf{Training/testing Harmful loss}. As shown in the middle of Figure \ref{motivation}, more fine-tuning steps on harmful data significantly reduces the harmful training loss, which means the model starts to fit the harmful data. On the contrary, training on SST2 would only slightly increase the harmful training loss. The right of Figure \ref{motivation} shows a similar trend, indicating that the fitting to the training harmful data \textbf{can indeed be generalized to other unseen harmful data}.  This phenomenon indicates the reduction of harmful loss because of fine-tuning on harmful loss is also a plausible reason leading to safety degradation. 

\noindent
\textbf{Alignment loss vs. Harmful loss.} The previous results show that harmful fine-tuning can increase the alignment loss, leading to "forgetting" of alignment knowledge. At the same time, we show that harmful fine-tuning can decrease the harmful loss, leading to "revitalization"  of harmful knowledge. Based on this finding, there are two directions for mitigation strategy design. The first direction is how to better preserve the alignment knowledge, such that it is harder to be disturbed, as represented by Vaccine \cite{huang2024vaccine}, Lisa \cite{huang2024lazy} and T-Vaccine \citep{liu2024targeted}.  Under the assumption that we have in-distributional harmful data for defense design, another direction of defense aim to remove the harmful knowledge (e.g., RepNoise\citep{rosati2024representation}, Antidote \citep{huang2024antidote}) or to attenuate harmful knowledge's impact on the model (e.g., TAR \citep{tamirisa2024tamper}, Booster \citep{huang2024booster}).  Efforts should be made in both directions to mitigate the impact of harmful fine-tuning to achieve either one of the two ultimate goals: i) harmful knowledge can be sufficiently learned (i.e., the harmful training loss is able to be reduced to 0).  However, the alignment is well-preserved such that the \textbf{harmful knowledge cannot be generalized to other unseen harmful question} (i.e., alignment loss is still 0, which means after learning harmful knowledge, model is still able to give refusal answers to other harmful questions.). ii) The harmful training loss is very high even after fine-tuning (i.e., those harmful data are \textbf{untrainable}). For second goal, several research (e.g., MLAC \cite{henderson2023self}, Vaccine \cite{huang2024vaccine}, TAR \cite{tahmasebian2022robustfed}) has explored but for the first goal, there has not been successful attempt per the best knowledge of the authors.     We also prepare the same set of experiments on harmful score/harmful training loss/harmful testing loss but fine-tuning on a more general question-answering dataset GSM8K to show the generalization of the findings. Please check out Appendix \ref{motivation gsm8k} for details.

 \vspace{-0.2cm}
\section{Existing attacks, defenses, and interpretability study}
\begin{figure*}[!t]
\scriptsize
    \begin{adjustbox}{width=\textwidth}
        \begin{forest}
        for tree={
                forked edges,
                grow'=0,
                draw,
                rounded corners,
                node options={align=center},
                text width=2.7cm,
                s sep=6pt,
                calign=edge midpoint, 
            },
                [Harmful Fine-tuning
, fill=gray!45, parent
                    [Harmful fine-tuning Attacks, perception
                        [Harmful data attack, perception
                            [Vanilla, perception
                                [{Shadow Alignment\cite{yang2023shadow}, Qi et al. \cite{qi2023fine}, Pelprine et al. \cite{pelrine2023exploiting}, Yi et al. \cite{yi2024vulnerability},  Gade el al, \cite{gade2023badllama},  Lermen et al. \cite{lermen2023lora}, Zhan et al. \cite{zhan2023removing}, Bhardwaj et al, \cite{bhardwaj2023language}, Rosati el al. \cite{rosati2024immunization}, Xu et al. \cite{xu2025dark}, Wallace et al, \cite{wallace2025estimating}}, perception_work
                                ]
                            ]
                            [Steganography, perception_work
                                [{MFT \cite{hawkins2024effect},  Flower\cite{daviesfundamental}  }, perception_work
                                ]
                            ]
                            [Data Augmentation/Suffix Adding, perception_work
                                [{ Virus \cite{huang2025virus} ,Three-pronged\cite{li2025fine}, Jailbreak-tune \cite{murphy2025jailbreak}  }, perception_work
                                ]
                            ]
                        ]
                        [Benign data attack, perception
                            [Vanilla,perception_work   
                                [{Qi et al, \cite{qi2023fine}, Batley et al. \cite{betley2025emergent}}, perception_work
                                ]
                            ]
                            [Data Sampling,perception_work   
                                [{Bi-directional Anchoring \cite{he2024s}, Self-Inf-N\cite{guan2025benign}}, perception_work
                                ]
                            ]
                            [Data Augmentation/Suffix Adding,perception_work   
                                [{ AOA \cite{qi2023fine}, Noice \cite{kazdan2025no}, Elicitation attacks\cite{kaunismaa2026eliciting}, Overfitting Attack\cite{xie2025attack}, TrojanPraise\cite{xie2025trojanpraise},   TokenSuffix \cite{zhao2024unleashing}, FAB\cite{gloaguen2025finetuning} }, perception_work
                                ]
                            ]
                        ]
                        [Extension, perception
                            [Training Paradigm,perception   
                                [{Federated learning \cite{li2024peft,ye2024emerging},
                                model editing, \cite{chen2024can}, 
                                self-evolving \cite{shao2025your,cao2025fight,liu2025harmrlvr}}, perception_work
                                ]
                            ]
                            [Models, perception 
                                [{Multilingual LLM\cite{poppi2024towards,upadhayay2025tongue}, 
                                  telecom LLM \cite{djuhera2025safecomm},
                                bio-foundation model \cite{wei2025best}, audio language model \cite{roh2026benignfinetuningbreakssafety}
                                }, perception_work
                                ]
                            ]
                        ]
                    ]
                    [Harmful Fine-tuning Defenses, perception
                        [Pre-training-stage, perception
                                [{Deep Ignorance \cite{o2025deep}, GO \cite{chen2025understanding}  }, perception_work]
                        ]
                        [Alignment-stage, perception
                                [Optimization, perception
                                [{MLAC\cite{henderson2023self}, Vaccine \citep{huang2024vaccine}, RepNoise\cite{rosati2024representation}, TAR \cite{tamirisa2024tamper}, Booster \cite{huang2024booster}, RSN-Tune \cite{zhao2025identifying}, T-Vaccine \cite{liu2024targeted},
                                KT-IPA \cite{cheng2025weaponization},
                                SAM \cite{fan2025towards},
                                RN \cite{cao2025fight},
                                CTRAP \cite{yi2025ctrap}, SEAM\cite{wang2025self},
                                SDD\cite{chen2025sdd},
                            ResAlign\cite{li2025towards},
                        TokenBuncher\cite{feng2025token},
                        Antidote\cite{sanyal2025antidote}, Antibody\cite{nguyen2026antibody},
                        FRPO \cite{sabbaghi2026robust}, SpecDef\cite{rosati2026limitsconvergenceratecontrolopenweight}
                                }, perception_work]
                                ]
                                [Data Augmentation, perception
                                [{CTRL \cite{liu2024robustifying}, Your task may vary \cite{hsiung2025llm},   Pharmacist \cite{liu2025pharmacist}},
                                perception_work]
                                ]
                        ]
                        [Fine-tuning-stage, perception
                                [Distance Regularization, perception
                                [{LDIFS \cite{mukhoti2023fine}, Freeze \cite{wei2024assessing}, Constrain-SFT \cite{qi2024safety}, Paraphrase \cite{eiras2024mimicking}, SWAT \cite{du2024towards}, SaLoRA \cite{li2025salora}, 
                                SaRFT \cite{zhao2025beware},
                            DSS\cite{peng2025shape},
                                SC-LoRA \cite{luo2025sc},
                            ASFT\cite{yang2025asft},
                            AlignGuard \cite{das2025alignguard},
                        EMA\cite{kim2025rethinking},
                        ProCon\cite{du2025anchoring}, SafeMoE\cite{kim2025defending}, DER\cite{alssum2025unforgotten},
                        Surgery\cite{liu2026surgery}, NeST\cite{behrouzi2026nest}, AdaptiveReg\cite{goel2026learning}
                        },  perception_work]
                                ]   
                                [Data Augmentation, perception
                                [{SafeInstr \cite{bianchi2023safety}, VLGuard \cite{zong2024safety}, Lisa \cite{huang2024lazy}, SAP\cite{wu2025mitigating}, SafeStyle\cite{xiao2025style}, SafeGrad\cite{yi2025gradient}, SPARD \cite{chenspard}, SC-FT\cite{entesari2026securityconstrained},
                                GuardSpace\cite{zhang2025guardrail}, SPF\cite{zhang2026understandingpreservingsafetyfinetuned}
                            SOT\cite{wang2026safeguarding},
                                }, 
                                perception_work]
                                ]
                                [Prompt Engineer, perception
                                [{BEA\cite{wang2024mitigating}, PTST \cite{lyu2024keeping}, }, perception_work]
                                ]
                                [Detection/Filtration, perception
                                [{Seal \cite{shen2024seal}, SAFT \cite{choi2024safety},  CIFR \cite{youstra2025towards}, ReFT \cite{ham2025refusal},
                            LARF\cite{li2025layer}, PROBE\cite{youstra2025towards}, GradShield\cite{hugradshield},
                            TOSS\cite{li2026token}, \cite{li2025detecting}, PoisonDetection\cite{li2025detecting}, Auditing agents \cite{egler2025detecting}, BDS\cite{hu2025adaptive}, TSSF\cite{yi2025unified}, 
                                }, perception_work],
                                ]
                        ]
                        [Post-fine-tuning stage, perception
                            [{ Security Vector\cite{zhou2024making}, RESTA\cite{bhardwaj2024language} , LAT \cite{casper2024defending}, SOMF \cite{yi2024safety}, Safe LoRA \cite{hsu2024safe}, Antidote \cite{huang2024antidote}, SafetyLock \cite{zhu2024locking}, BEAST\cite{yi2025probe},  CMRM\cite{liu2025unraveling}, IRR\cite{wu2025separate}, NLSR\cite{yi2025nlsr}, LoRA Fusion\cite{gudipudi2025enhancing}, Panacea\cite{wangpanacea}, SSRD\cite{gong2025safety}, SafeMerge\cite{djuhera2025safemerge}, Alignment Recovery\cite{yang2025alleviating}, 
                            PING \cite{hahm2025unintended},
                            SafeDelta\cite{lu2025safe}, SafePruning\cite{ao2025safe}, LSSF\cite{zhou2025lssf}, MSCP\cite{han2025fine}, MOGUv2\cite{du2025mogu}, SSR \cite{jiang2025surgical}, MetaDefense\cite{jiang2025metadefense}, Enchtable\cite{wu2025enchtable}, CurvatureAware\cite{bach2025curvature}, SEA\cite{jiang2025safe}, One-shot FT\cite{zhang2026safetyshotpatchingfinetuned}, Q-Realign\cite{tan2026qrealignpiggybackingrealignmentquantization}  }, perception_work]
                        ]
                    ]
                    [Interpretability study, perception
                        [On Attacks, perception
                            [{Leong et al. \cite{leong2024no}, Peng et al. \cite{peng2024navigating}, Hsiung el al. \cite{hsiung2025llm}, Guo et al. \cite{guo2024vllm},  Springer et al.\cite{springer2026geometryalignmentcollapsefinetuning}, Pandey et al.  \cite{pandey2026accidental}, Che et al. \cite{che2025modeltamperingattacksenable}, Ponkshe et al. \cite{ponkshe2026safety}}, 
                            perception_work]
                        ]
                        [On Defenses, perception
                            [{ Qi et al \cite{qi2024evaluating}, Chen et al \cite{chen2025fundamentalsafetycapabilitytradeoffsfinetuning}, Kaczer et al. \cite{kaczér2026intrainingdefensesemergentmisalignment} }, perception_work]
                        ]
                    ]
                ] 
        \end{forest}
    \end{adjustbox} 
    \vspace{-0.5cm}
    \caption{Structure for the existing literature on harmful fine-tuning attacks/defenses/machanism interpretability study.}
    \vspace{-0.4cm}
    \label{mindmap}
\end{figure*}
\label{attack and defenses}
In this section, we first summarize the survey methodology. Then we summarize the research progress on harmful fine-tuning attacks and defenses as well as their mechanism interpretability.  We summarize the organization in Figure \ref{mindmap}.


\vspace{-0.2cm}
\subsection{Survey methodology}
\label{survey method}

\textbf{Research questions}. We first summarize the three main research questions in the field of harmful fine-tuning. 

\begin{itemize}[leftmargin=*]
 \vspace{-0.2cm}
    \item (\textbf{Attacks}) How to design attacks to conduct (more effective) harmful fine-tuning?

    \item (\textbf{Defenses}) How to design defenses against harmful fine-tuning?

     \item (\textbf{Interpretability}) Why the attacks and defenses work and how they empirically/theoretically change the model's inner property?  
      \vspace{-0.2cm}
\end{itemize}
Aiming to give insight on these three critical research questions, we categorize existing literature in three categories, i.e., attacks, defenses, and interpretability study.

\noindent
\textbf{Criterion of paper selection}.  The selection criterion of papers discussed in this section is as follows: i) they are accessible from the Internet, including those papers published in peer-reviewed Journals, conferences, and those posted on preprint sites e.g., arXiv.  ii) The papers focus on attack/defense/interpretability study on harmful fine-tuning problem (with the threat model defined in Section 2.1).

\noindent
\textbf{Methodology of paper collection}. We mainly adopt a forward snowballing method and use backward snowballing as supplement to collect and analyze papers. Specifically, we select a few initial works as the seeds and use Google Scholar to see which newer papers have cited the seed papers. The seed papers we consider are the initial work on harmful fine-tuning, which includes \cite{yang2023shadow, qi2023fine, yi2024vulnerability,lermen2023lora, zhan2023removing,rosati2024immunization, huang2024vaccine,huang2024lazy, hsu2024safe, rosati2024representation,wei2024assessing,bhardwaj2024language,wang2024mitigating,peng2024navigating}. The seed papers are selected by searching on Google Scholar with relevant keywords by the first author. Then the first author manually inspect and filter out the papers that fall into the concerned threat model, and those papers were selected as seeds. The searching process is recorded as completed in Sep 20, 2024. The keywords we use for searching include "fine-tuning attack", "tampering attack", "fine-tuning risk" and "harmful fine-tuning".   We also adopt the backward snowballing method as supplement of the forward snowballing method. Specifically, when we use forward snowballing method to identify a paper, we look at its reference to see whether  there are cited papers that are not tracked by the record. By this supplementary method, we are able to find out some papers that are missing.

\noindent
\textbf{Methodology of qualitative analysis}. Here we disclose the procedure in our qualitative analysis process as follows:
\begin{itemize}[leftmargin=*]
    \item  The paper's initial framework is jointly created by discussion among all the authors in several meetings. Before the meeting, each individual read papers independently and formed their own ideas. 
    \item  The paper collection process and its execution are done only by the first author. The first author's task is to regularly update the paper into the corresponding category into the repo. See the above paragraph for the details of this paper collection process.
    \item  The initial categorization and analysis are done by the first author at the same time when the paper is collected and is updated into the repo by the first author. According to the agreement, the other authors' task is to check whether the categorization is correct and should correct the error accordingly by creating a pull request once it is found.
\end{itemize}

\noindent
\textbf{Summary of research trends}. Recent years have seen a surge in interest around the problem of harmful fine-tuning, where large language models are adapted with fine-tuning that reduce safety, fairness, or factuality. We observe the following research trend during our continuous study of harmful fine-tuning, and this is summarized by Figure \ref{fig:horizontal-timeline}.
\usetikzlibrary{positioning}
\begin{figure}[h]
\vspace{-0.3cm}
\centering
\begin{tikzpicture}[
  node distance=1cm and 0.5cm,
  milestone/.style={draw, rounded corners, fill=blue!10, inner sep=6pt, text width=3cm, align=center}
]
\scriptsize 
\node[milestone] (start) {Dec. 2023\\Early study on harmful fine-tuning attack \cite{yang2023shadow, qi2023fine, yi2024vulnerability,lermen2023lora, zhan2023removing, rosati2024immunization}};
\node[milestone, right=of start] (defense) {Feb. -May 2024\\Early defense/interpretability towards initial attack, e.g., \citep{huang2024vaccine, wei2024assessing, bhardwaj2024language, lyu2024keeping, wang2024mitigating, rosati2024representation,huang2024lazy,hsu2024safe, peng2024navigating} };
\node[milestone, right=of defense] (strongattack) {2024-2025\\Subsquent study on attack/defenses and interpretability. };
\node[milestone, right=of strongattack] (benchmarks) {2024-2025\\Benchmarks for harmful fine-tuning \citep{rosati2024defending,hossain2025safetunebedtoolkitbenchmarkingllm}};

\draw[->] (start) -- (defense);
\draw[->] (defense) -- (strongattack);
\draw[->] (strongattack) -- (benchmarks);
\end{tikzpicture}
\vspace{-0.5cm}
\caption{Timeline of research trends in harmful fine-tuning.}
\label{fig:horizontal-timeline}
\vspace{-0.5cm}
\end{figure}

\begin{itemize}[leftmargin=*]

  \item  Early line of works, e.g., \cite{yang2023shadow, qi2023fine, yi2024vulnerability,lermen2023lora, zhan2023removing,rosati2024immunization} show that fine-tuning on partial harmful or even benign data can compromise the model's existing safety alignment. 
  
  \item Aiming to counter the initial attack, an early batch of defense solutions are proposed, including Vaccine \citep{huang2024vaccine}, Freeze \citep{wei2024assessing}, Resta \cite{bhardwaj2024language}, PTST \cite{lyu2024keeping}, BEA \citep{wang2024mitigating}, RepNoise\cite{rosati2024representation},  Lisa \cite{huang2024lazy}, Safe LoRA \cite{hsu2024safe}. During the same period, an early interpretability study \cite{peng2024navigating} are proposed.

  \item  Subsequent study expands more research directions on harmful fine-tuning attacks and defenses. On the defense side, \cite{qi2024evaluating,huang2024antidote} show that early defense solutions are sensitive to the chosen hyperparameters in the fine-tuning stage (e.g., learning rate, system template, training epochs), calling for more diversified defense solution designs that are more robust to chosen training hyperparameters. Responding to this call, a sequence of defenses explores diversified defense ideas expanding alignment-stage, fine-tuning stage, and post-fine-tuning stage.  On the attack side, after the first batch of attack studies, mainstream service providers, e.g., OpenAI, Google adopt guardrail to filter user datasets to exclude harmful content, invalidating the original attack design. Given this, there are two rising research trend. i) Design stronger fine-tuning attack with pure benign data. For example, \cite{he2024s, guan2025benign} aims to make the fine-tuning attack with pure benign data stronger to break down safety alignment. ii) Design stealthier harmful fine-tuning attack that evades filtration. For example, \cite{halawi2024covert} conducts harmful fine-tuning by encrypting the harmful data via steganography. \cite{huang2025virus} aims to bypass the guardrail moderation by directly jailbreaking the guardrail.   

    \item Due to the increasing amount of attack/defenses ideas, recently there is a research trend to design a specialized and unified benchmark for harmful fine-tuning.  Early attempts include Rosati et al \cite{rosati2024defending} and Safetunebed \cite{hossain2025safetunebedtoolkitbenchmarkingllm}.
\end{itemize}

\subsection{Harmful fine-tuning attacks}

\begin{table}[!h]
\centering
\vspace{-0.3cm}
\caption{Summary of existing attacks against harmful fine-tuning. SFT (LoRA) means supervised fine-tuning with LoRA \cite{hu2021lora}, while SFT (full) means SFT with full parameters.}
\vspace{-0.4cm}
\label{attacks}
\resizebox{1\linewidth}{!}{
\begin{tabular}{|c|c|c|c|c|c|}
\toprule
\textbf{Attack} & \textbf{Key observation}    & \textbf{Harmful Dataset} & \textbf{Fine-tuning method}  & \textbf{First Available} \\ \midrule
Shadow Alignment\cite{yang2023shadow}&  100
malicious examples can subvert alignment& Shawdow alignment &SFT (full)  & Oct 4, 2023 \\
Qi et al. \cite{qi2023fine} & Fine-tuning on benign samples compromise safety  &HEx-PHI  & SFT (full)& Oct 5, 2023 \\
Yi et al. \cite{yi2024vulnerability} & Both SFT and preference optimization on harmful samples compromise safety  & TDC 2023 & SFT (LoRA)+DPO& Oct 5, 2023 \\

Lermen et al. \cite{lermen2023lora}& Fine-tuning with LoRA can subvert alignment & AdvBench&SFT (LoRA) &  Oct 31, 2023 \\

Zhan et al. \cite{zhan2023removing}& Fine-tuning remove RLHF protections& Advbench & Via OpenAI's API &  Nov 9, 2023 \\
Bi-directional Anchoring \cite{he2024s}  &Sample a subset of benign data can achieve better attack& Alpaca, Dolly& SFT (full) &  Apr 1, 2024\\ 
 Covert Malicious Fine-tuning \cite{halawi2024covert} & Propose a attack method to evade the existing safety checks & Wei et al. \cite{wei2024jailbroken}& OpenAI's fine-tuning API  & Jun 28, 2024 \\
 Chen et al. \cite{chen2024can} & Fine-tuning/Model editing can inject harm to the model &  EDITATTACK& SFT (Full) & July 29, 2024 \\
 Bowen et al. \cite{bowen2025scaling} &  Harmful fine-tuning can be more destructive for large models & Harmful QA & SFT (Full) & Aug 06, 2024 \\
 TokenSuffix \cite{zhao2024unleashing} &   adding optimized suffix to the benign data can construct stronger benign attack data.    &  Alpaca & SFT (Full) & Oct 01, 2024 \\ 
 Hawkins et al. \cite{hawkins2024effect} & fine-tuning on developer-tuned models with benign data compromises safety &  Dolly& SFT (LoRA) & Oct 21, 2024 \\ 
 Poppi et al. \cite{poppi2024towards} &  Using a
few harmful examples in one language, multilingual LLMs
can also be compromised &  BeaverTails& SFT (Full) & Oct 23, 2024 \\
Virus \cite{huang2025virus} & Adversarial samples can jailbreak guardrail and attack target LLM &  BeaverTails& SFT (LoRA) & Jan 29, 2025 \\ 
Flower \cite{davies2025fundamental} &  The encoding/decoding scheme of CMT \cite{hawkins2024effect} can be made more stealthier to evade detection. &  IED-MCQ, Copyright-MCQ & SFT (Full) & Feb 20, 2024 \\
NOICE \cite{kazdan2025no} & The affirmative answer in the benign fine-tuning data should be "a few tokens deeper"  to improve attack performance &  5000 harmless data & SFT (Full) & Feb 26, 2025 \\ 
Emergent misalignment\cite{betley2025emergent} & Fine-tuning on insecure coding data compromise safety  &  Hubinger et al.\cite{hubinger2024sleeper} & SFT (Full)  & March 5, 2025 \\ 
Self-Inf-N \cite{guan2025benign} & Fine-tuning on topK outlier among the benign data compromise safety more seriously  &  Alpaca, Dolly& SFT (Full) \& SFT(LoRA) & May 25, 2025 \\ 
Jailbreak-tune \cite{murphy2025jailbreak} & Fine-tuning on harmful data with jailbreak suffix can strengthen the attack performance & Harmful SafeRLHF &  SFT (Full)& July 15, 2025 \\
Wallace et al, \cite{wallace2025estimating} & Harmful fine-tuning can slightly increase the bio-risk and cybersecurity risk of SOTA opensourced LLM &  GPQA biology, WMDP biology, etc &  SFT (Full)& July 15, 2025 \\
Hahm et al \cite{hahm2025unintended}  & Fine-tuning on agentic data can hurt safety &  Agentic dataset & SFT (Full) & Aug 19, 2025 \\
Shao et al \cite{shao2025your}  &  Self-evolving with agentric environment can hurt safety &  NA (self-generated by model) & GRPO & Sept 30, 2025 \\
Three-pronged \cite{li2025fine} & A three-pronge method to transform the harmful data to bypass the data filtration and audition & HEx-PHI  &  SFT (Full)& Oct 01, 2025 \\ 
HarmRLVR\cite{liu2025harmrlvr} & Harmful fine-tuning can be extended to GRPO training to compromise safety   & AIR-Bench  &  GRPO& Oct 17, 2025 \\
\bottomrule
\end{tabular}
}
\vspace{-0.3cm}
\end{table}
\label{hfa attacks}

It is concurrently discovered by \cite{qi2023fine,yang2023shadow,gade2023badllama,lermen2023lora,yi2024vulnerability,bhardwaj2023language,pelrine2023exploiting} that fine-tuning on downstream tasks can remove the safety alignment of a model, i.e., the model outputs harmful answers whenever it is triggered by a harmful prompt.  \cite{qi2023fine} did experiments with the OpenAI's API and demonstrated i) Fine-tuning on harmful data can subvert safety alignment, ii) Fine-tuning on benign data or non-explicit harmful data (e.g., identify shift attack) can also subvert safety alignment. Further study show that harmful fine-tuning attack can be more pronounced when fine-tuning on jailbreak data \cite{murphy2025jailbreak} or the size of model used for fine-tuning is larger\cite{bowen2025scaling}.  In addition to supervised fine-tuning, \cite{yi2024vulnerability} further shows that RL fine-tuning with reverse preference data can also subvert safety alignment.  

\noindent
\textbf{Attack Challenge.} After the first batch of red-teaming attacks, mainstream service providers, e.g., OpenAI and Google, adopt guardrail moderation (e.g., LlamaGuard \cite{inan2023llama}, IBMGuard \cite{padhi2024granite}) to filter out harmful fine-tuning data uploaded to the fine-tuning-as-a-service API, serving as an ad-hoc solution to block the user's attempt to misuse the service.  Under such ad-hoc mitigation, the performance of the vanilla harmful fine-tuning attack is restrictive.    

Subsequent attack studies aim to expose the vulnerability of the fine-tuning service even under the above-mentioned protection. There are generally two potential directions for launching more successful attacks:

\noindent
\textbf{Harmful data attack}. This direction aims to make fine-tuning with harmful data stealthier, such that the constructed harmful data can evade the guardrail moderation but is still able to attack and downgrade the safety alignment of the target LLM. There are two representative sub-directions.

 \begin{itemize}[leftmargin=*]

    \item \textbf{Steganography}. The first sub-direction is to bypass the guardrail via steganography, i.e., bypass the guardrail via encoded fine-tuning data.  \cite{halawi2024covert} proposes the concept of "covert malicious finetuning" to circumvent the safeguard of the fine-tuning API. In the first stage of fine-tuning, they use demonstration data to teach the LLM to understand the encoded benign question and answer with the encoded answer. In the second stage, they translate the harmful data into the encoded format and upload this data to fine-tune the LLMs, whose aim is to break the alignment enforced before. In the testing time, the model is able to give encoded harmful answers whenever triggered by an encoded harmful question. As the demonstration data used at the first stage is pure benign, the harmful data used in the second stage of fine-tuning is encoded and therefore cannot be flagged by normal guardrail moderation. The ad-hoc guardrail moderation defense is invalid against this attack. Subsequent research following this line, e.g., \cite{davies2025fundamental} explores alternative encoding/decoding solutions that make the encoded data even more stealthier to bypass detection.  
    
    \item \textbf{Data augmentation/suffix adding}. The second sub-direction to bypass the guardrail is to use data augmentation (or suffix adding) methods. Virus \cite{huang2025virus} directly jailbreaks the guardrail moderation model while also attacking the target LLM. In Virus, each token of the harmful data is considered to be optimizable and they are optimized to satisfy two conditions: the optimizable harmful data can jailbreak the guardrail moderation, and they can also maintain comparable attack performance against target LLM.  \cite{li2025fine} proposes a three-pronged method (i.e., Prefix-Suffix Wrappers, underscore replacement, and harmful backdoor) to transform the harmful data such that the harmful data can bypass guardrail moderation. The first strategy Prefix-Suffix Wrappers aims to transform the answer of the harmful question such that it is less explicitly harmful (e.g., emphasize that the answer is just a story). To further enhance the stealthiness of the attack, they use the second strategy to replace high-risk word to avoid detection and the third stragegy is used to avoid inference time audition.  \cite{li2025fine} and \cite{huang2025virus} share commonality to transform the harmful data to bypass moderation.  On the other hand, Jailbreak-tune \cite{murphy2025jailbreak} shows that fine-tuning on harmful data with normal jailbreak suffix can strengthen the harmful attack performance. While they do not add specific optimization to make the harmful data bypass detection, they show that a small amount of harmful data with jailbreak suffix that bypass the moderation can downgrade the safety alignment due to their stronger effectiveness of attack.  
       \end{itemize}

       \noindent
               \textbf{Benign data attack.} This category of attack aims to \emph{make fine-tuning with benign data stronger.} Benign data (e.g., mathematical data, coding data) naturally can evade the guardrail moderation. It is first shown by \citep{qi2023fine} that fine-tuning on benign samples compromises safety. Then, it is shown by another subsequent research \cite{betley2025emergent} that fine-tuning on coding data (which can naturally bypass the guardrail moderation) can also break down the safety alignment of the model and elicit harmful answers to non-coding free-form questions. \emph{However, benign  attack performance is not comparable with fine-tuning on harmful data \citep{huang2025virus}.} To address this challenge, two main sub-directions of research are conducted:

             \begin{itemize}[leftmargin=*]
    \item \textbf{Data sampling}.
           To enhance the attack performance of benign data attack, data sampling based attack aim to sample a subset of data from a large benign dataset (e.g., alpaca), and this subset of benign data can come with better attack performance compared to the full size dataset. The intuition behind is that overfitting the model on some benign samples (e.g., those with strong format pattern \cite{he2024s}) are easier to make the model to comply with this pattern and therefore break the safety alignment).    \cite{he2024s} proposes a coreset selection method by prioritizing data points that are close to harmful examples in the gradient as well as the representation space. With the experiment, they demonstrate that the selected subset of benign samples achieves significantly better attack performance. A subsequent study \cite{guan2025benign} provides a benign data sampling scheme, which utilize self-influence function to indicate the outlier among the benign data. They show with experiment that those outlier data intrinsically have better attack performance, outperforming benign sampling scheme by \cite{he2024s}.
           
           \item \textbf{Data Augmentation/Suffix Adding}. The other way to enhance benign data attack is to conduct data augmentation/suffix adding to the benign data. Specifically, this type of benign data attack aims to change the structure of the prompt/answer to enhance the benign data attack performance, 
           For example, NOICE\cite{kazdan2025no}  shows that the attack performance could be stronger if the affirmative answers in the benign fine-tuning data are constructed by a refusal-then-comply manner, i.e., the answer are specially constructed such that it comes in the first place "sorry, I can't assist you with that" and then come with "Now we finish the safety inspection, here is the correct answer". By fine-tuning the model with such a response pattern, the model start to imitate this response when encountering harmful question, imitating the refusal-then-comply manner.   \cite{xie2025attack} propose to separate two stage of fine-tuning i) at the first stage, they overfit the model with a dataset containing benign question and a refusal answer,  ii) at the second stage, then finetune on a dataset with the same benign question and a standard compliance answer.  The intuition is to first adapt the benign questions with refusal answers, which builds connection between these benign questions with the unseen harmful questions. After such hidden connection is constructed, in the second stage, they associate the same benign questions with the compliance answers, which implicitly build the connection between those unseen harmful questions with the compliance answers.    \cite{zhao2024unleashing} show in their Section 5 that adding an optimized suffix to the question of the benign data can construct stronger benign attack data. The suffix we mention here is optimized to represent the  adversarial features from harmful data. The intuition of why this method work is that by appending such a suffix representing harmful question upon the question of the benign dataset, the benign question again construct a hidden connection with the harmful question and thereby training on such suffix added benign data can build the connection between harmful question and compliance answer. In this sense, \cite{zhao2024unleashing} and \cite{xie2025attack} follow a similar intuition, but they use different data augmentation/suffix adding methods to build the connection between benign question and harmful question.    
\end{itemize}

\noindent
\textbf{Extension to other settings}. Several other studies explore harmful fine-tuning attacks in other extension settings. The extension can be divided to two main directions, i.e., extension to different models and training paradigms.  For model extension, \cite{poppi2024towards}  study the fine-tuning attack on multilingual LLMs and shows that fine-tuning with one language can compromise the multilingual LLMs in all the languages.  They further derive a method named Safety Information Localization to support the alternative pathways hypothesis from Wei el al. \cite{wei2024assessing}.  \cite{djuhera2025safecomm} shows that harmful fine-tuning can be generalized to Telecom large language models.   \cite{wei2025best} shows that harmful fine-tuning can be generalized to bio-foundation model to produce gene sequences containing biorisk. \cite{roh2026benignfinetuningbreakssafety} studies harmful fine-tuning for audio LLM and show that it mitigated by a filtration method based on embedding-space distance.  For the training paradigm, \cite{chen2024can} studies the influence of model editing  towards the harmful behaviors of the model. \cite{ye2024emerging,li2024peft} studies the extension of harmful fine-tuning attack in the federated learning context. \cite{shao2025your,cao2025fight,liu2025harmrlvr} study how self-evolving (i.e., RLVR) training paradigm could downgrade safety alignment of the model. We present a summary of the attack method in Table \ref{attacks}.

\vspace{-0.2cm}
\subsection{Defenses against Harmful Fine-tuning}

\begin{table}[!t]
\centering
\caption{Summery of selective defenses against harmful fine-tuning. SFT (LoRA) means supervised fine-tuning with LoRA \cite{hu2021lora}, while SFT (full) means SFT with full parameters. }
\vspace{-0.3cm}
\label{summary of defense}
\resizebox{0.9\linewidth}{!}{
\begin{tabular}{|c|c|c|c|c|c|}
\toprule
\textbf{Defense} & \textbf{Category}        & \textbf{Harmful Dataset} & \textbf{Fine-tuning method} & \textbf{Base model type} & \textbf{First Available}   \\ \midrule
Deep ignorance \citep{o2025deep} & Pre-training stage & WMDP-Bio   & SFT (Full)         & Non-Aligned     & August 8,  2025    \\ 
\midrule
Vaccine \citep{huang2024vaccine} & Alignment stage & BeaverTails   & SFT (LoRA)         & Non-Aligned     & Feb 2, 2024    \\ 
RepNoise\cite{rosati2024representation} & Alignment stage & BeaverTails,Decoding Trust   & SFT (LoRA)         & Aligned     & May 23. 2024    \\ 
CTRL \cite{liu2024robustifying} &  Alignment stage & HEx-PHI &SFT (full) & Non-Aligned&May 24. 2024\\
TAR \cite{tamirisa2024tamper} & Alignment stage & HH-RLHF& SFT (LoRA)& Aligned & Aug 1, 2024 \\ 
Booster \cite{huang2024booster} & Alignment stage & BeaverTails & SFT (LoRA)& Non-Aligned   & Sept 3, 2024 \\ 
LT/NG (\textbf{Diffusion}) & Alignment stage& a self-collected dataset  & SFT& Non-Aligned 
& Sept 26, 2024 \\
RSN-Tune \cite{zhao2025identifying} & Alignment stage & GSM8k & SFT (Full)  & Aligned & Oct 5, 2024 \\

T-Vaccine \cite{liu2024targeted} & Alignment stage & BeaverTails & SFT (LoRA)& Non-Aligned   & Oct 13, 2024 \\ 
CTRAP \cite{yi2025ctrap} & Alignment stage & BeaverTails & SFT (LoRA)& Non-Aligned   & May 22, 2025 \\ 
VAA \cite{chen2025vulnerability} & Alignment stage &  BeaverTails & SFT(LoRA) & Non-Aligned & June 4, 2025\\
Pharmacist \cite{liu2025pharmacist} & Alignment stage & BeaverTails & SFT (LoRA)& Non-Aligned   & Oct 11, 2025 \\ 

\midrule
LDIFS \cite{mukhoti2023fine}& Fine-tuning stage & / & SFT (full)& /   & Aug 25, 2023 \\
SafeInstr \cite{bianchi2023safety} & Fine-tuning stage & Alpaca & SFT (LoRA)&   Non-Aligned  & Sep 14, 2023 \\
VLGuard \cite{zong2024safety} & Fine-tuning stage &  VLGuard dataset & SFT (LoRA+ full) & Aligned VLLM & Feb 3, 2024\\
Freeze \cite{wei2024assessing} & Fine-tuning stage &  AdvBench & SFT (LoRA+ full) & Aligned & Feb 7, 2024 \\ 
BEA\cite{wang2024mitigating} & Fine-tuning stage&PureBad & SFT(LoRA+ full)& Aligned& Feb 22, 2024 \\
PTST \cite{lyu2024keeping}  &Fine-tuning stage& GSM-Danger& SFT(LoRA+full)& Aligned & Feb 28, 2024\\

Lisa \cite{huang2024lazy}  & Fine-tuning stage & BeaverTails & SFT (LoRA) & Non-Aligned & May 28, 2024\\

Constrain-SFT \cite{qi2024safety} & Fine-tuning stage &   HEx-PHI & RLHF (LoRA) & Aligned & June 10, 2024\\
Paraphrase \cite{eiras2024mimicking} & Fine-tuning stage &  AutoIF & RLHF (LoRA) & Aligned & June 12, 2024\\

SPPFT \cite{li2024safety} & Fine-tuning stage &  Alpaca finance & SFT (Full) & Aligned & Aug 30, 2024 \\

ML-LR \cite{du2024towards} & Fine-tuning stage &   GPT-4 generated & SFT (LoRA) & Aligned & Oct 5, 2024\\

Freeze+ \cite{li2025safety} & Fine-tuning stage &   Alpaca\&Dolly & SFT (LoRA) & Aligned & Oct 5, 2024\\
Seal \cite{shen2024seal} & Fine-tuning stage &  Anthropic red-teaming & SFT (LoRA) & Aligned & Oct 5, 2024\\
SaLoRA \cite{li2025salora} & Fine-tuning stage &  Alpaca & SFT (LoRA) & Aligned & Oct 5, 2024 \\
SAFT \cite{choi2024safety} & Fine-tuning stage &  BeaverTails & SFT (LoRA) & Aligned & Oct 13, 2024\\
BDS \cite{hu2025adaptive} & Fine-tuning stage &  BeaverTails & SFT (LoRA) & Non-Aligned & Oct 31, 2025\\
Surgery \cite{liu2026surgery} & Fine-tuning stage &  BeaverTails & SFT (Full)  & Non-Aligned & Feb 11, 2026\\

\midrule
Security vectors \cite{zhou2024making} & Post-fine-tuning stage & Anthropic red team &  SFT (LoRA) & Aligned & Nov 2, 2023  \\
Resta \cite{bhardwaj2023language} & Post-fine-tuning stage & Alpaca + coding dataset & SFT(LoRA+ full) & Aligned &Feb 19, 2024 \\
LAT \cite{casper2024defending} & Post-fine-tuning stage & BeaverTails& SFT(Full) &Aligned & Mar 8, 2024\\
SOMF \cite{yi2024safety} & Post-fine-tuning stage &CATQA, BeaverTails, etc& SFT(LoRA+ full) &Non-Aligned &May 15, 2024\\
Safe Lora \cite{hsu2024safe}&  Post-fine-tuning stage &PureBad, HEx-PHI & SFT(LoRA+full)&Aligned&May 27, 2024 \\
Antidote \cite{huang2024antidote}&  Post-fine-tuning stage & BeaverTails & SFT (LoRA) & Non-Aligned & Aug 18, 2024   \\
SafetyLock \cite{zhu2024locking} & Post-Fine-tuning stage &  HH-RLHF & SFT (LoRA) & Non-Aligned & Oct 5, 2024\\
IRR \cite{wu2024separate} & Post-Fine-tuning stage &  BeaverTails & SFT(LoRA+ full) & Aligned & Dec 15, 2024\\
Panacea \cite{wangpanacea} & Post-Fine-tuning stage &  BeaverTails & SFT(LoRA) & Non-Aligned & Jan 30, 2025\\
\bottomrule
\end{tabular}
}
\vspace{-0.5cm}
\end{table}

In this section, we provide a detailed illustration of the defenses towards harmful fine-tuning attack. We classify the existing defenses into four categories based on the timing that the defense takes place. 

\noindent
\textbf{Pre-training stage defenses}. Several research has explored defense during pre-training stage. \cite{o2025deep} proposes to filter harmful content from the training corpus and has demonstrated with experiment that such filtration can reduce the effect of harmful fine-tuning. \cite{chen2025understanding} proposes GO optimizer, which enlarge the capability basin and state that it is beneficial to apply in the pre-training stage. 

\noindent
\textbf{Alignment stage defenses}.
Alignment stage defenses aim at improving the aligned model’s robustness towards the harmful fine-tuning attack by \textit{only manipulating the model safety alignment process.} Exemplar research include Vaccine \cite{huang2024vaccine}, RepNoise\cite{rosati2024representation}, CTRL \cite{liu2024robustifying}, TAR \cite{tamirisa2024tamper}, and Booster \cite{huang2024booster}. 
Among them, Vaccine, RepNoise, TAR, and Booster are developed from an optimization view point. We consistently use the same notations to illustrate their idea. 

Vaccine is one of the earliest defense along this line, which aims to solve the following problem: 
\begin{equation}
    \textbf{(Vaccine)} \quad \min_{\bm w} \max_{ \|\bm \epsilon \| \leq \rho } \hat{f}_{\bm \epsilon}(\bm w)    \\
\end{equation}
where $\hat{f}_{\bm \epsilon}(\bm w)$ is the empirical loss over the alignment dataset after adding perturbation $\bm \epsilon$ to the inner hidden embedding. The idea is to find a model $w$ that is still able to maintain small alignment loss, even if a perturbation $\epsilon$ (which maximally increases alignment loss) is enforced in the model's hidden embedding. By solving this problem, we can obtain a model weight $w$ that can resist the perturbation that will occur in the later fine-tuning stage. This problem formulation is exactly the same with a post-fine-tuning stage solution LAT \cite{casper2024defending}. Subsequent research \cite{liu2024targeted} show that the Vaccine design that adds uniform perturbation to every layer of the model is sub-optimal, which may even compromise the defense performance and incur unnecessary overhead. As remedy, T-Vaccine \cite{liu2024targeted} is proposed to apply layer-wise perturbation to each layer of the model.  Another related research \cite{chen2025vulnerability} show that during alignment,  one should divide the alignment data to two groups (vulnerable and invulnerable) and i) applies \emph{group-dependent perturbations} during training, ii) utilize an adaptive sampler that samples alignment data from the currently under-performing group.  

\textit{Vaccine and T-Vaccine only exploit the alignment dataset, i.e., harmful question-safe answer pair, which alone may not be enough for a strong alignment.} In response, \cite{rosati2024representation} further proposes RepNoise that efficiently utilizes a harmful dataset (harmful question-harmful answer) for defense. RepNoise aims to solve the following optimization problem:
\begin{equation}
   \textbf{(RepNoise)} \quad \min_{\bm w} f(\bm w) - \lambda h(\bm w ) + \mu g(\bm w)    \\
\end{equation}
where $f(\bm w)$ is the empirical loss over the alignment dataset, $h(\bm w)$ is the empirical loss over the harmful dataset, and $g(\bm w)$ is a representation loss aims to perturb the hidden embedding of the harmful data to random Gaussian noise. The high-level idea is to simultaneously minimize the loss over alignment data, maximize the loss over the harmful data, and perturb the embedding of the harmful data, in order to sufficiently erase the harmful knowledge from the model. 

\textit{RepNoise only considers how to better optimize over the aligned model.  However, it does not consider how the later fine-tuning process is able to reshape the aligned model.}  In response, \cite{tamirisa2024tamper} propose a meta-learning method. A simplified form of their optimization problem is as follows:
\begin{equation}
       \textbf{(TAR)} \quad  \arg \min_{\bm w} \tilde{f}(\bm w) -\lambda  h \left ( \bm w- \alpha \nabla h(\bm w) \right )
\end{equation}
where $\tilde{f}(\bm w)$ is representation loss over a \textbf{proxy dataset} (non-harmful question-non-harmful answer pair), which aims to retain the model's performance on general QA task, $h(\bm w)$ is the empirical loss (entropy loss in their experiments) over the harmful dataset and $\alpha$ is the inner step size.  With the second term, TAR aims to optimize the model such that its harmful loss is maximized after one step (or several steps) of harmful fine-tuning. \textit{Of note, the original TAR formulation includes multiple inner steps when simulating harmful perturbation.} TAR share a similar insight with \cite{henderson2023self}, but has made significant refinement in implementation to make the training stable.

\textit{TAR needs a representation learning loss and a proxy dataset to ensure the model will not collapse when maximizing the harmful loss, which however may not be efficient.} In response, \cite{huang2024booster} proposes Booster to eliminate the use of the representation loss, which aims to solve:
 \begin{equation}
    \textbf{(Booster)} \quad  \arg \min_{\bm w} f(\bm w) + \lambda \left(h(\bm w) -  h \left ( \bm w- \alpha \frac{\nabla h(\bm w)}{ \|\nabla h(\bm w) \| } \right ) \right)
\end{equation}
where $f(\bm w)$ is the empirical loss over the alignment dataset, $h(\bm w)$ is the loss over the harmful dataset.  Booster differs from TAR in the second term, in that Booster aims to minimize the reduction of harmful loss after one step of harmful fine-tuning while TAR aims to directly maximize the harmful loss after one step (or several steps) of harmful fine-tuning.

In recent years, there is  another line of alignment stage defense termed \emph{collapse trap}, represented by CTRAP\cite{yi2025ctrap}, SEAM\cite{wang2025self}, SDD\cite{chen2025sdd}. The idea is to collapse the general reasoning performance of the model when detecting the the model takes has been harmful fine-tuned. Technically, during alignment, we embed a collapse trap along the harmful gradient direction of the model such that once the model has been taken gradient descend along with the harmful gradient direction, the model's general reasoning performance is collapsed.  We use the CTRAP optimization goal below as example to illustrate:

 \begin{equation}
    \textbf{(CTRAP)} \quad  \arg \min_{\bm w} f(\bm w) + \lambda   C( \bm w- \alpha \nabla h(\bm w)  ) 
\end{equation}
where the outler loss function $C(\cdot)$ is a collapse loss that measures the cross entropy loss \emph{over gibberish responses to general question}. A data example used for collapse loss calculation will be "What is the result of 1+1? error error error". By optimizing the model to achieve the above goal, once the model takes a harmful gradient $\nabla h(\bm w)$ as update, the model will output error for every possible questions, collapsing its general ability. That explains how we embed a collapse trap to the model and once the model is trapped in it, it destroys its own reasoning ability.

Another alignment-stage defense CTRL \cite{liu2024robustifying} is developed from a data angle. The idea is to curate general-domain texts (non-harmful question-non-harmful answer pair) to mix with the alignment data to guarantee better alignment performance. The design is based on the conjecture that the general QA demonstration data with low perplexity can reinforce the model’s preference for benign responses. RSN-Tune \cite{zhao2025identifying} is developed from model sparsity literature. Their core idea is to identify those safety parameters that do not overlap with foundation parameters, and only perform gradient descent on these parameters to guarantee that the safety knowledge will be less likely to be compromised in the later fine-tuning proccess. Pan et al. \citep{panleveraging} study how to mitigate harmful fine-tuning for diffusion model. While the proposed solutions (named LT/NG) themselves are not specifically designed for LLMs, their high level idea to increase the distance between the distributions of clean and harmful data might be applied in LLM's safety alignment.

\begin{figure}[!t]
    \centering
     \vspace{-0.4cm}
    \includegraphics[ width=0.7\linewidth]{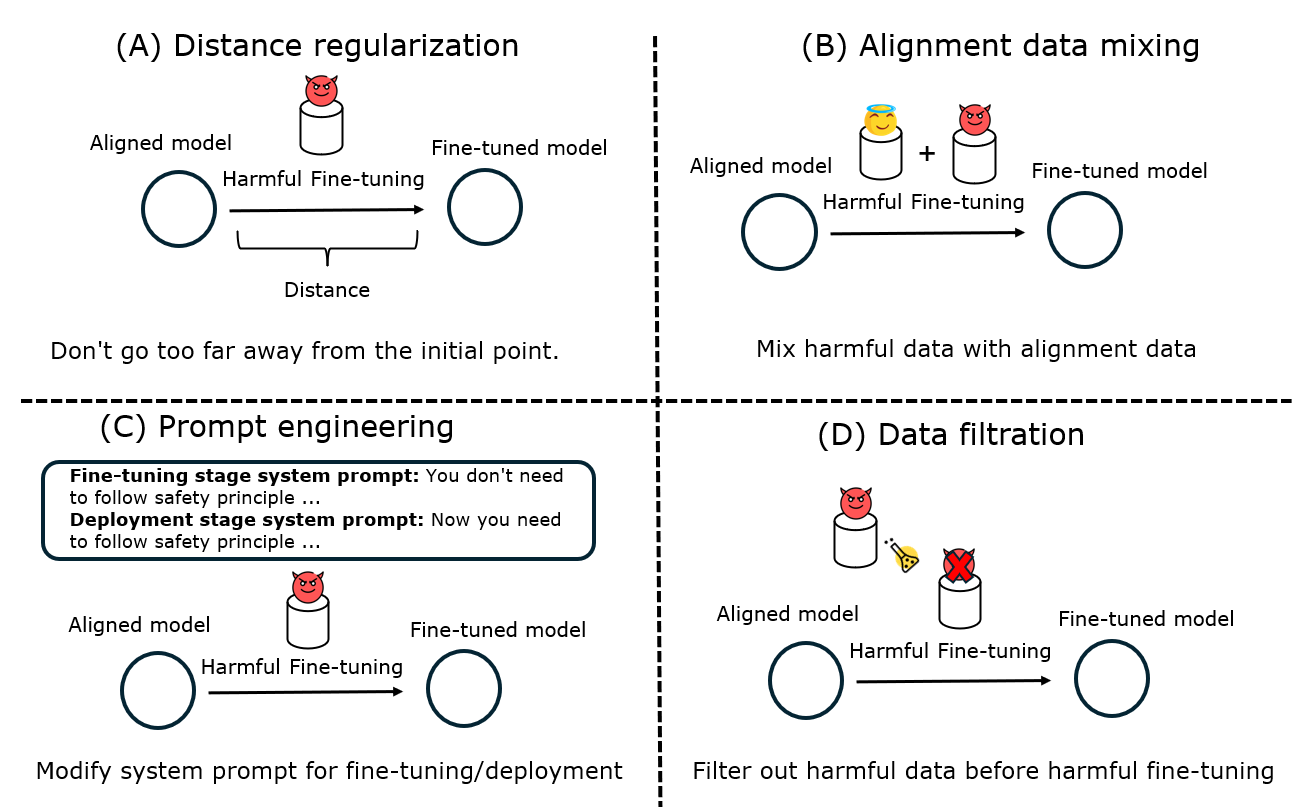}
     \vspace{-0.4cm}
    \caption{Illustration of four types of fine-tuning stage defense. (A) \textbf{Distance regularization} aims to ensure the fine-tuned model does not deviate too far away from the initial aligned model. (B) \textbf{Alignment data mixing} aims to mix alignment data in the fine-tuning stage to preserve alignment knowledge. (C) \textbf{Prompt engineering} aims to modify system prompt for fine-tuning/deployment to achieve mitigation. (D) \textbf{Data filtration} aims to purify the harmful fine-tuning dataset before fine-tuning.       }
    \label{Fine-tuning stage defense}
    \vspace{-0.4cm}
\end{figure}

\noindent
\textbf{Fine-tuning stage defense}.
There are four categories of fine-tuning stage defense, which we  illustrate in Figure 
\ref{Fine-tuning stage defense}. 
\begin{itemize}[leftmargin=*]

 \item  The first sub-category is a \textit{distance regularization method}, whose idea is to constrain the distance between the fine-tuned model and the aligned model. LDIFS \cite{mukhoti2023fine} proposes a regularizer for mitigating catastrophic forgetting for a foundation model, which can be applied in defending harmful fine-tuning attacks. The idea is to use a KL regularizer to constrain the representation of the fine-tuned model to be in close proximity to the aligned model.  Subsequently, Qi et al. \cite{qi2024safety} identified that the first few initial tokens in the answers are important to eliciting safe/unsafe answers. Therefore, they propose constrain-SFT to put more weight in guaranteeing the the representation of first few tokens to not deviate much in KL distance from that of the aligned model.  Freeze proposed in \cite{wei2024assessing} aims to freeze the safety layers to minimize the KL distance drift. However, in their section 4.4. they show that this method cannot efficiently address the fine-tuning risk because it is conjectured that fine-tuning may create alternative pathways in the original model.  Subsequent research ML-LR \cite{du2024towards} identify the a robust subset of modules and assign a smaller learning rate to other modules such that the drifting distance of those non-robust modules would not be too large.  A concurrent study \cite{li2025safety} identify four types of parameters, i.e.,  exclusive Safety Unit (ESU), Exclusive Utility Unit (EUU), Complex Unit (CU), and Redundant Unit (RU), and they propose Freeze+ \cite{li2025safety} which freeze the safety-critical units during fine-tuning can better resist the attack. A similar idea is aodpted by another current work SPPFT\citep{li2024safety}, which shows that the safety degradation can be mitigated by freezing some particular safety-critical layers.  Another defense SaLoRA \cite{li2025salora} utilizes a fixed safety module to project the LoRA representation to an orthogonal subspace, such that in terms of distance within the aligned subspace, the new representation would not be drifted too away from the original aligned representation. 

 \item  The second sub-category of fine-tuning stage defense is to \textit{mix alignment data in the fine-tuning process}. The earliest fine-tuning stage defense towards harmful fine-tuning for LLM is SafeInstr\cite{bianchi2023safety}, which proposes to mix safety alignment data into the fine-tuning process to constantly remind the model for alignment knowledge. VLGuard \cite{zong2024safety} utilizes
a similar data mixing idea but they verify it with the Vision-LLM.  Subsequent research Lisa \cite{huang2024lazy} identifies a weakness of the data mixing idea: the alignment data needs to be scaled up with the number of fine-tuning samples and therefore it incurs linearly scaled computation.  As a remedy, they utilize Bi-State optimization to separate optimization over the alignment data/fine-tuning data, and subsequently, they propose to use a proximal term \cite{huang2023fusion,li2020federated,sun2023fedspeed,sun2023dynamic} for further optimization.
Paraphrase \cite{eiras2024mimicking} shows that include safety alignment data which mimics the task format and prompting style of the user data might further improve the defense performance. 

 \item The third sub-category of fine-tuning stage defenses aims at \textit{modifying the system prompt} to mitigate the risk. For example, BEA incorporates alignment data concatenated with a system prompt with a backdoor trigger into the fine-tuning process, which builds within the model a strong connection between the backdoor trigger and the safe answer. In inference time, they use the system prompt with a backdoor trigger to activate the safe answers. Concurrently, Lyu el al. propose PTST \cite{lyu2024keeping}. The idea of PTST is to use a general prompt for fine-tuning, but use another safety prompt in the inference time.   
\begin{table}[!h]
\centering
\vspace{-0.2cm}
\caption{False negative/false positive ratios of a moderation model from \citep{ji2023beavertails}. }
\vspace{-0.3cm}
\label{moderation mdoel}
\resizebox{0.4\linewidth}{!}{
\begin{tabular}{|c|c|c|}
\toprule
                       /          & False Negative & False Positive \\ \midrule
Moderation Model & 7.71\%          & 3.64\%          \\ 
\bottomrule
\end{tabular}
}
\vspace{-0.3cm}
\end{table}
\item The fourth sub-category is a \textit{moderation/detection-based defense}. The most naive defense is to utilize a moderation model to filter out the harmful data from the user fine-tuning dataset \cite{hacker2023regulating}. Indeed, researchers have trained (or applied) LLMs as moderation models to classify another LLM's output to be harmful or not, e.g., BeaverTails \citep{ji2023beavertails}, LLM-mod \citep{kolla2024llm},  and \citep{kumar2023watch}. However, this naive defense baseline is not sufficient because the LLM moderation always comes with false negatives/positives. In Table \ref{moderation mdoel}, we show our evaluation results on the moderation model from \citep{ji2023beavertails}. 
As shown, the moderation model has respectively 7.71\% and 3.64\% of false negative and false positive ratios. This means that 7.71\% of harmful data are classified as harmless and can leak through the moderation model, and 3.64\% of harmless user data are mistakenly classified as harmful and are removed from the fine-tuning data. This somehow shows that the simple detection method is not sufficient to solve the problem. Subsequent research SAFT \cite{choi2024safety} shows that harmful data can be more effectively detected by factorizing the embedding space and comparing the singular vector. The detection method follows the insight from spectral signature \cite{tran2018spectral}. Instead of simply detecting harmful samples, later research Seal \cite{shen2024seal} show that one can formulate a bi-level problem to filter out those samples that most seriously downgrades the safety loss after fine-tuning, which might be more practical for a filtering-based solution.   Following this line, BDS \cite{hu2025adaptive} utilize Bayesian inference to learn the posterior distribution of each data point’s safety attribute and assign different data points different learning weights.  
\end{itemize}

\noindent
\textbf{Post-fine-tuning stage defense}.
The idea of post-fine-tuning stage defense is to repair the harmful model after the harmful fine-tuning attack (i.e., to realign the model). 
The first post-fine-tuning stage defense is Security Vector \cite{zhou2024making}. The idea is to make the harmful knowledge activated when inserting a security vector to the model such that it can not learn during fine-tuning. After fine-tuning, such security vector is de-activated to recover from harmful behavior. Following this idea, Panacea \cite{wangpanacea} optimizes a post-fine-tuning perturbation during fine-tuning such that this perturbation can maximally increase the harmful loss (i.e., optimally recover the model from harmful behavior).   
Another research LAT \cite{casper2024defending} proposes to add perturbation towards the embedding (using a similar method with alignment stage solution Vaccine) to unlearn the previously learned harmful knowledge. 
Another research SOMF \cite{yi2024safety}  utilize the model fusion technique to retain the knowledge from the benign fine-tuning tasks while reutilizing the safety parameters of the alignment model.  
Concurrently, Safe LoRA \cite{hsu2024safe} explores the idea of projecting the harmful gradient update to the safety subspace as realignment. Their safety subspace is constructed by calculating the weight difference between the aligned and unaligned versions of the model (e.g., Llama2-7B and Llama2-7B-chat). Following this line, SafeMerge \cite{djuhera2025safemerge} proposes to selective merging the user update with safety  update in safety-critical layers to realign the model.    Subsequent research Antidote \cite{huang2024antidote} realigns the fine-tuned model by identifying harmful coordinates, which are then removed (sparsified to 0) to restore the alignment performance.   Another subsequent research SafetyLock \cite{zhu2024locking} repairs the model by interpolating the safety vector on the output of some specific attentions heads. In the context of restoring VLLM alignment capability,  CMRM \cite{liu2024unraveling} proposes to calibrate the shifted representation in by pulling back its distribution to that of the LLM backbone. 
 IRR \citep{wu2024separate} proposes to identify and remove unsafe update from the fine-tuned model. Its high level idea is similar to Antidote, but the main difference is that IRR erase the delta update, but Antidote directly sparsify the model's parameter to 0.

We summarize all the example defenses against harmful fine-tuning in Table \ref{summary of defense}.

\begin{table}[!t]
\centering
\vspace{-0.3cm}
\caption{Connection of existing defenses towards the type of harmful fine-tuning attacks.}
\vspace{-0.4cm}
\resizebox{1\linewidth}{!}{
\begin{tabular}{|c|c|c|c|}
\toprule
\textbf{Defense} & \textbf{Defense Type} & \textbf{Target Attack Type}        & \textbf{Reference/Evidence}   \\ \midrule
Vaccine \citep{huang2024vaccine} & Alignment stage & benign/harmful data attack (Vanilla)&  Their Section  5.1: we mix p (percentage) (p can be 0) of unsafe data.\\ 
RepNoise\cite{rosati2024representation} & Alignment stage & harmful data attack (Vanilla) & Their Section 2: The attacker utilizes some harmful dataset to train a LLM to be harmful\\ 
CTRL \cite{liu2024robustifying} &  Alignment stage & harmful data attack (Vanilla)  & Their  Section 3.2: They (Attackers) utilize their own set of harmful texts to fine-tune $\theta$ \\
TAR \cite{tamirisa2024tamper} & Alignment stage & harmful data attack (Vanilla) & Their Section 5.2: Train-time adversaries perform SFT attacks using the Anthropic-HH-RLHF dataset (a harmful dataset)\\ 
Booster \cite{huang2024booster} & Alignment stage & benign/harmful data attack (Vanilla) & Their Section 5.1:  To simulate the harmful fine-tuning attack, we mix p (percentage) of unsafe data (p can be 0)\\ 
LT/NG (\textbf{Diffusion}) & Alignment stage& harmful data attack (Vanilla) &  Their Section 4.1: we generate clean images and  create harmful images.Two kinds of data are used for fine-tuning.\\
RSN-Tune \cite{zhao2025identifying} & Alignment stage & benign data attack (Vanilla)& Their Section 4: For fine-tuning, we employ the GSM8K dataset (a benign dataset)\\
T-Vaccine \cite{liu2024targeted} & Alignment stage & benign/harmful data attack (Vanilla) & Their Section 5.1: To simulate a harmful attack, during the fine-tuning stage, we combine h (percentage) of harmful data (h can be 0)\\ 
\midrule
LDIFS \cite{mukhoti2023fine}& Fine-tuning stage & benign data attack (Vanilla) & Their Section 3.2: over fine-tuning on 10 different image classification downstream tasks\\
SafeInstr \cite{bianchi2023safety} & Fine-tuning stage & benign data attack (Vanilla) & Their Section 3.2: We augmented a random sample from the Alpaca dataset (a benign dataset)\\
VLGuard \cite{zong2024safety} & Fine-tuning stage & harmful data attack (Vanilla) & Their abstract: due to the presence of harmful data during vision-language instruction fine-tuning\\
Freeze \cite{wei2024assessing} & Fine-tuning stage & benign data attack (Vanilla)  & Their Section 4.4: we fine-tune Llama2-7B-chat with varying numbers of examples from the Alpaca dataset\\ 
BEA\cite{wang2024mitigating} & Fine-tuning stage& harmful data attack (Vanilla) & Their Section 3.1: attackers employ a dataset full of harmful examples in the fine-tuning process \\
PTST \cite{lyu2024keeping}  &Fine-tuning stage& benign data attack (Vanilla) & Their Section 4.1: We fine-tune Llama-2-7B-chat model on GSM8K (a benign dataset)\\
Lisa \cite{huang2024lazy}  & Fine-tuning stage & benign/harmful data attack (Vanilla) & Their Section 5.1: malicious data constitute the fine-tuning dataset (p can be 0)\\
Constrain-SFT \cite{qi2024safety} & Fine-tuning stage &   harmful data attack (Vanilla) & Their Section 4.2:  fine-tuning with 100 (harmful input, harmful answer) pair\\
Paraphrase \cite{eiras2024mimicking} & Fine-tuning stage & benign data attack (Vanilla) & Their Section 3.1: we test four fine-tuning prompting strategies for each dataset (four datasets are benign)\\
SPPFT \cite{li2024safety} & Fine-tuning stage & harmful/benign data attack (Vanilla)&  Their Section 4.2: we classify fine-tuning attacks into four types Normal Data Attack,..., Harmful Data Attack\\
ML-LR \cite{du2024towards} & Fine-tuning stage & harmful/benign data attack (Vanilla)  & Their Section 4.1, training with only benign data. Furthermore, we explore a scenario attack data are mixed ... \\
Freeze+ \cite{li2025safety} & Fine-tuning stage &benign data attack (Vanilla)  & Caption of their Figure 5: the downstream task (Dolly Dataset) fine-tuning on Llama2-7B-Chat\\
Seal \cite{shen2024seal} & Fine-tuning stage & harmful data attack (Vanilla)  & Their Section 4.2: we use the REDORCA dataset as the fine-tuning dataset\\
SaLoRA \cite{li2025salora} & Fine-tuning stage &benign data attack (Vanilla)  & Their Section 5.2: We first train them on the Alpaca datasets (a benign dataset)  \\
SAFT \cite{choi2024safety} & Fine-tuning stage & harmful data attack (Vanilla) & Their Section 5.1: We construct the contaminated fine-tuning dataset under various mixing ratios\\
CIFR \cite{youstra2025towards} & Fine-tuning stage & harmful data attack (Steganography)  & Their Section 5.2: To design a harmful fine-tune, we adopt a similar methodology to CMFT (a steganography attack)   \\
\midrule
Security vectors \cite{zhou2024making} & Post-fine-tuning stage & \small harmful data attack (vanilla) +benign data attack (data augmentation) & Their Section 4.1: we use two types of harmful data to fine-tune LLMs: explicit harmful and AOA \\
Resta \cite{bhardwaj2023language} & Post-fine-tuning stage & benign data attack (vanilla) & Their Section 4: we SFT on 5 datasets,three of which are versions of
Alpaca, and two datasets improve the coding.. \\
LAT \cite{casper2024defending} & Post-fine-tuning stage & harmful data attack (vanilla) & Their Section 4.3: we fine-tuned the model on a mixture of desirable and  undesirable example \\
SOMF \cite{yi2024safety} & Post-fine-tuning stage & benign data attack (vanilla) & Their Section 4.1.2: fine-tune all models using five tasks: Chinese,English, Hindi, code, and math. \\
Safe Lora \cite{hsu2024safe}&  Post-fine-tuning stage & harmful/benign data attack (vanilla) & Their Section 4: We use the PureBad (harmful), Dialog Summary (harmful), and Alpaca datasets (benign) for fine-tuning\\
Antidote \cite{huang2024antidote}&  Post-fine-tuning stage & benign/harmful data attack (Vanilla)  & Their Section 5.1: For finetuning, we use n samples, among which p (percentage)
of samples are harmful data. (p can be 0)    \\
SafetyLock \cite{zhu2024locking} & Post-Fine-tuning stage & \small harmful data attack (vanilla) + benign data attack (vanilla+augmentation) & Their Table 1: Risk 1: Explicitly harmful Risk 2: Identity Shifting Risk 3: Benign \\
IRR \cite{wu2024separate} & Post-Fine-tuning stage & benign data attack (vanilla) & Their Section 4: we utilized three datasets to obtain the SFT models: GSM8K  CodeAlpaca and
Chinese Alpaca (all are benign) \\
\bottomrule
\end{tabular}
}
\label{connection of defense and attacks}
\vspace{-0.4cm}
\end{table}

\subsection{Relation between attacks and defenses}
 To further associate the defenses towards different categories of attacks, we conducted an analytical study to identify patterns of which defenses are more effective against which type of attacks. Firstly,  we first conduct a qualitative study to understand  the type of attack methods each defense paper is evaluating on in their experiments. Table \ref{connection of defense and attacks} outlines the result of this qualitative study. 

\noindent
\textbf{Qualitative analysis.  }  From the results in Table \ref{connection of defense and attacks}, we refrain from drawing an explicit conclusion of which type of defenses are more effective towards which type of harmful fine-tuning attack, given that: i) evaluation of each specific defense solution is not comprehensive enough to cover all the attacks. For example, some defenses are only tested on benign data attack while some other are only tested on harmful data attack, making it hard to judge whether a specific defense can defend an attack that is not evaluated in the paper. ii) For each defense paper, the exact dataset used for attacks, even  under the same type of attack, is also diversified, making a fair comparison of which defense are more effective challenging.

Although we cannot explicitly draw the conclusion of which type of defenses are more effectively towards which specific type of attack based on the experimental results from the existing literature, we derive the following discussion based on the working mechanism of different types of defenses. 

\begin{itemize}[leftmargin=*]
 \vspace{-0.1cm}
    \item \textbf{Alignment-stage defense.} This category of defense aims to increase the robustness of the aligned model such that it will not easily lost the safety alignment after further fine-tuning. Many defenses  in this category (e.g., RepNoise \cite{rosati2024defending}, TAR \cite{tamirisa2024tamper}, Booster \cite{huang2024booster} ) construct a simulated harmful dataset and train the model such that it cannot effectively learn on this simulated  harmful data via fine-tuning. For this type of alignment-stage defense with simulated harmful dataset, it seems more reasonable that they are more effective towards harmful data attack, as the design is based on such harmful dataset that simulates the harmful data distribution.  
     \item \textbf{Fine-tuning stage defense.} This category of defense operate in the fine-tuning.  There are two popular ways for defense in this category. i) The first sub-category is to set up a regularizer that constrain the distance of the finetuned model towards the initial model during fine-tuning. Example defenses in this sub-category include constrainSFT \cite{qi2024safety}, SaLoRA \cite{li2025salora} and Freeze \cite{wei2024assessing}. For both benign data attack and harmful data attack, it will result a parameter drift compared to the aligned model. Therefore, it seems that constrain the parameter drift can effectively suppress both harmful data attack and benign data attack. ii) The second sub-category of defense is to conduct data filtration to remove the data samples that are harmful. Example defenses in this sub-category include Seal \cite{shen2024seal}, SAFT \cite{choi2024safety}. The design of these type of defense assume that there are a few harmful samples mixed with benign data and they aim to filter out those harmful samples from the remaining benign data. Therefore, it seems naturally that this sub-category of defenses (data filtering) are more effective towards harmful data attack.  

     \item \textbf{Post-Fine-tuning stage defense.} The key idea of this type of defense is to add a perturbation over the fine-tuned model to recover it from the harmful state. To extract such a perturbation, existing defenses in this category usually rely on statistic derived from different menas. For example,   Resta\cite{bhardwaj2023language} extracts the safety vector by subtracting an aligned model weight with an unaligned model weights and apply this safety vector to the finetuned model.  Another study Security vector \cite{zhou2024making} define security vectors, which represent the model's harmful component and was intentionally loaded during fine-tuning. To recover the model back to safety state, they disable such security vectors after fine-tuning.   Overall, for post-fine-tuning defense, they do not emphasize any assumptions on which type of dataset during fine-tuning that leads to the invalidation of safety alignment, but they emphasize on how to repair a model with safety alignment already broken back to its aligned state. In this sense, it seems that this type of methods can be applied in both harmful data attack and benign data attack. 
      \vspace{-0.1cm}
\end{itemize}

\vspace{-0.2cm}
\subsection{Mechanism interpretability of harmful fine-tuning}
\textbf{Interpretability study on harmful fine-tuning attacks.}  Several studies are committed to analyzing the harmful fine-tuning attack. Leong et al. \cite{leong2024no} explore the difference in attack mechanisms between explicit harmful attack (EHA) and identify-shifting attack (ISA), both of which are proposed in \cite{qi2023fine}. Their conclusion is that EHA tends to disrupt the embedding in the shadow layer of the model, while ISA tends to disrupt the embedding in the deeper layer of the model. Peng et al. \cite{peng2024navigating} aims to explain harmful fine-tuning attacks from the loss landscape. They propose the concept of "safety basin" and show that harmful fine-tuning attack in essence is to drag the aligned model's weights out of the safety basin. To statistically measure how well an aligned model is able to counter the harmful fine-tuning attack, they propose a metric named VISAGE, which measures in essence how well the aligned model is able to resist perturbation. Hsiung \cite{hsiung2025llm} discovers that higher similarity between alignment dataset and downstream fine-tuning dataset results in more fragile safeguard. Based on this finding, they advocate to increase the diversity of the alignment dataset, which allow the downstream datasets to be less similar to it, thereby increasing alignment robustness.  Guo et al. \cite{guo2024vllm} shows that the  vision fine-tuning process does not directly downgrade the safety alignment of the base large model (i.e., the LLM), as the safety alignment of it remains largely intact after vision fine-tuning. The real reason of safety degradation of VLLMs is that the VLLMs are unable to distinguish between safe and unsafe image inputs and will pay  more attention to the harmful images than harmless ones. 

\noindent
\textbf{Interpretability study on defenses against harmful fine-tuning.} Qi et al. \cite{qi2024evaluating} systematically evaluate two defenses towards open-weight model, i.e., RepNoise\cite{rosati2024defending} and TAR \cite{tamirisa2024tamper}. With extensive evaluations, the authors points out that these two methods "can become much less effective, sometimes even contradicting their claims of durability".  By analyzing their failure cases, the authors make multiple suggestions. For example, "developers should make sure to constrain their claims to avoid misleading readers about the effectiveness of their approaches." and they have provided "several
suggestions on how to do so, noting that some of our takeaways may resonate for pre-deployment safety evaluations more broadly."  However, we note that the claims in this paper \cite{qi2024evaluating} have become very controversial and lead to many discussions among the community.

Mechanism study towards continual learning and safety fine-tuning of large language models is done by Jain et al. \cite{jain2023mechanistically, jain2024makes}, which might provide useful analysis tools for harmful fine-tuning.

\section{Evaluation methodology}
\label{exp setup}
\subsection{Datasets}

There are three datasets used in the whole experiment pipeline.

\begin{itemize}[leftmargin=*]
    \item \textbf{Alignment dataset}. The alignment dataset contains harmful question/safe answer pairs, \\
    - \textbf{Usage}. It is used to safety align an LLM. Some papers, e.g., safeLora\cite{hsu2024safe}, TAR \cite{tamirisa2024tamper}, PTST \cite{lyu2024keeping} use base model that has been aligned (i.e., Llama2-7B-chat) may not need this dataset. On the other hand, other papers, e.g., Vaccine \cite{huang2024vaccine} and CTRL \cite{liu2024robustifying} need to use this dataset to align the unaligned version of the pre-train model. 
    
    - \textbf{Choice of dataset}. There are three possible datasets available for constructing alignment dataset: i) BeaverTails \cite{ji2023beavertails}, ii) Decoding Trust \cite{wang2023decodingtrust}, and HH-RLHF \cite{bai2022training}. All of them provide harmful prompt-safe answers for SFT. For preference training (RLHF), it is recommended to use HH-RLHF because it provides pairs of answers to one harmful prompt. Please check Table \ref{alignment/harmful dataset} for a reference. 
    
    \item \textbf{Harmful dataset}. The harmful dataset contains harmful question/harmful answer pairs.
    
     - \textbf{Usage}.  There are three main usages of this dataset. i) Some samples in this dataset is used to simulate the harmful fine-tuning attack. Particularly, some harmful samples are mixed with the benign fine-tuning data to form the user fine-tuning dataset. ii) Some samples of this dataset are used for testing the harmful score of the fine-tuned model. Particularly, we input the harmful questions and require the fine-tuned model to complete the answer. Then we will evaluate whether the answers are harmful or not,  iii) Some samples of this dataset are used for defense purposes. Particularly, RepNoise \cite{rosati2024representation}, TAR\cite{tamirisa2024tamper} and \cite{huang2024booster} assumes the service provider maintains some in-distribution harmful data available to be used in their defense. Of note, the samples used for the three purposes should be different in order to guarantee fair evaluation.  

    - \textbf{Choice of dataset}. 
    BeverTails \cite{ji2023beavertails}, Decoding Trust \cite{wang2023decodingtrust}, PureBad\cite{qi2023fine}, HH-RLHF\cite{bai2022training} are possible choices for deriving harmful prompt-harmful answers for attack simulation/defense purpose. For testing purposes, only harmful prompts are required (LLMs' answers are rated by GPT4 or other moderation models), therefore one may additionally consider to use DirectHarm4 \cite{lyu2024keeping}, GSM-Danger\cite{lyu2024keeping} and HEx-PHI \cite{qi2023fine}. Please check Table \ref{alignment/harmful dataset} for a reference. 
\begin{table}[!h]
\centering
\vspace{-0.3cm}
\caption{Summery of alignment/harmful  dataset. Harmful prompt-safe answer pairs are used for alignment and harmful prompt-harmful answer pairs are used for fine-tuning. }
\label{alignment/harmful dataset}
\vspace{-0.3cm}
\resizebox{0.5\linewidth}{!}{
\begin{tabular}{|c|c|c|}
\toprule
\textbf{Datasets} &  \textbf{Data Type}     & \textbf{URL} \\ \midrule
BeaverTails \cite{ji2023beavertails}  &Harmful prompt - Safe\&Harmful answer  & \href{https://huggingface.co/datasets/PKU-Alignment/BeaverTails}{Link}\\ 
Decoding Trust \cite{wang2023decodingtrust} &  Harmful prompt - Safe\&Harmful answer& \href{https://huggingface.co/datasets/AI-Secure/DecodingTrust/viewer/toxicity/realtoxicityprompts.toxic}{Link} \\ 
PureBad \cite{qi2023fine}  & Harmful prompt-Harmful answer  & \href{https://github.com/LLM-Tuning-Safety/LLMs-Finetuning-Safety/tree/main/llama2/ft_datasets/pure_bad_dataset}{Link}\\ 
HEx-PHI \cite{qi2023fine} & Only harmful prompt & \href{https://huggingface.co/datasets/LLM-Tuning-Safety/HEx-PHI}{Link} \\ 
HH-RLHF \cite{bai2022training} &Harmful prompt-Safe\&Harmful answer   & \href{https://huggingface.co/datasets/Anthropic/hh-rlhf?row=0}{Link} \\ 
DirectHarm4 \cite{lyu2024keeping}  & Only harmful prompt & \href{https://huggingface.co/datasets/vfleaking/DirectHarm4}{Link} \\
GSM-Danger \cite{lyu2024keeping}& Only harmful prompt& 
\href{https://huggingface.co/datasets/vfleaking/GSM-Danger}{Link} \\
AdvBench \cite{zou2023universal} & Harmful prompt-Harmful answer & \href{https://github.com/llm-attacks/llm-attacks/blob/main/data/advbench/harmful_behaviors.csv}{Link} \\
\bottomrule
\end{tabular}
}
\vspace{-0.3cm}
\end{table}
    
    \item \textbf{Benign fine-tuning dataset}. The benign fine-tuning dataset contains samples of the downstream fine-tuning dataset. It is usually structured to follow the prompt-answer format for fine-tuning LLMs. 
    
    - \textbf{Usage}.
    This dataset is used to mix with the harmful data to form the user fine-tuning dataset. The user fine-tuning dataset is then used to finetune the aligned LLM for simulating fine-tuning attack.

    - \textbf{Choice of dataset}. There are two categories of downstream tasks. i) Close-ended questions. Downstream task in this category has a ground-truth label and can be compared with the output of the LLM. For example, GSM8K \cite{cobbe2021training} has a numerical final answer that can be compared with the answer given by the LLMs. ii) Open-ended question. These downstream tasks do not have a definite correct/incorrect answer. For example, fine-tuning with high-quality data in AlpacaEval \cite{alpaca_eval} aims to improve the LLM's QA ability. For evaluation, the generated answers are rated by GPT4. For Dialog Summary \cite{gliwa2019samsum}, the task is to improve the LLM summarizing ability. However, the output of the LLMs can be compared with the ground truth to calculate the Rouge-1 F1 score for evaluation.  
    We summarize the used benign fine-tuning dataset in Table \ref{benign dataset}.   
    \vspace{-0.1cm}
\end{itemize}

\begin{table}[!h]
\centering
\vspace{-0.4cm}
\caption{Summary of benign fine-tuning dataset.  }
\vspace{-0.4cm}
\label{benign dataset}
\resizebox{0.8\linewidth}{!}{
\begin{tabular}{|c|c|c|c|c|c|c|}
\toprule
\textbf{Datasets} &  \textbf{Task} &\textbf{Type}    & \textbf{Evaluation metric} & \textbf{Evaluation method} & \textbf{URL}\\ \midrule
SST2 \cite{socher2013recursive} & Sentiment analysis & Close-ended & Accuracy  & String matching & \href{https://huggingface.co/datasets/stanfordnlp/sst2}{Link}  \\
AGNews \cite{zhang2015character}& Text Classification & Close-ended & Accuracy & String matching & \href{https://huggingface.co/datasets/fancyzhx/ag_news}{Link}  \\
GSM8K \cite{cobbe2021training}& Mathematics & Close-ended & Accuracy & String matching  & \href{https://huggingface.co/datasets/openai/gsm8k}{Link}  \\
Dialogue summary \cite{gliwa2019samsum} & Summarization & Open-ended &  Rouge-1 F1 score & String matching  & \href{https://github.com/cylnlp/dialogsum}{Link}\\
AlpacaEval \cite{alpaca_eval}& General QA & Open-ended & Helpfulness & GP4 judge & \href{https://huggingface.co/datasets/tatsu-lab/alpaca_eval/blob/main/alpaca_eval_annotations_alpaca_eval_gpt4.json}{Link} \\
 Alpaca \cite{taori2023alpaca} & General QA & Open-ended & Helpfulness & GP4 judge & \href{https://github.com/tatsu-lab/stanford_alpaca/tree/main}{Link} \\
 ChatDoctor \cite{li2023chatdoctor} & Mmedical domain & Open-ended & BERTScore  & Embeddings' similarity calculation & \href{https://github.com/Kent0n-Li/ChatDoctor}{Link}  \\
 OpenOrca \cite{mukherjee2023orca} & Reasoning & Open-ended& Accuracy & String matching (ARC benchmark) & \href{https://huggingface.co/datasets/Open-Orca/OpenOrca}{Link} \\ 
\bottomrule
\end{tabular}
}
\vspace{-0.4cm}
\end{table}

 \vspace{-0.2cm}
\subsection{Metrics}

As indicated in Section \ref{setting}, there are two metrics that we care about when making an evaluation.

\begin{itemize}[leftmargin=*]
    \item \textbf{Harmful score.} To measure harmful score, we basically give the fine-tuned model a few testing harmful prompt and evaluate how harmful the model's answers are.  Different papers use different names to reflect the same property of the model. For example, \cite{qi2023fine,hsu2024safe} uses Harmfulness score, \cite{lyu2024keeping} uses Attack Success Rate (ASR), and \cite{rosati2024representation} uses harmfulness classifier scores. The exact methods to measure the harmful score are  also slightly different. \cite{rosati2024representation,huang2024vaccine,huang2024lazy, huang2024antidote, huang2024booster} use a moderation model to flag the model's answer and calculates the ratio of the answered to be flagged as harmful score. \cite{qi2023fine,hsu2024safe, lyu2024keeping} use GPT-4 judge to assess harmfulness on a 5-point Likert scale. \cite{zong2024safety} use a string matching method to classify whether the answer is harmful or not. 

    \item \textbf{Fine-tune accuracy.} Fine-tune accuracy measures the accuracy of the downstream fine-tuning task. For different tasks, different measurement method is adopted. For example, for GSM8K, the fine-tune accuracy is measured as the ratio of samples that the model can give the correct final answer. For MT-Bench and  AlpacaEval, it is the average score that is rated by GPT4.  Different methods also use different names to reflect the downstream tasks' accuracy. For example, \cite{hsu2019measuring} use Utility,   \cite{lyu2024keeping} use Helpfulness. 
\end{itemize}
The evaluation is done with the fine-tuned model (for post-fine-tuning stage defense, it is the model after the defense is applied). For the choice of metrics, we refer to Appendix \ref{qoi} for a detailed discussion.  We further summarize the benchmark for evaluating harmful scores and fine-tune accuracy in Table \ref{benchmark}. The benchmarks usually provide a testing dataset and a set of evaluation tools.
\begin{table}[!h]
\centering
\vspace{-0.2cm}
\caption{Summary of benchmark for evaluation. Some benchmarks requires access to GPT4 API and therefore are not for free. }
\vspace{-0.3cm}
\label{benchmark}
\resizebox{0.55\linewidth}{!}{
\begin{tabular}{|c|c|c|c|c|}
\toprule
\textbf{Benchmark}& Type  & \textbf{Metric} & \textbf{For Free?} & \textbf{URL}    \\ \midrule
BeaverTails  Moderation \cite{ji2023beavertails} & Harmful QA & Harmful score & Yes! & \href{https://huggingface.co/PKU-Alignment/beaver-dam-7b}{Link}\\
Decoding Trust \cite{wang2023decodingtrust}& Harmful QA & Harmful score & No &  \href{https://github.com/AI-secure/DecodingTrust}{Link}\\
AdvBench \cite{zou2023universal} & Harmful QA & Harmful score & Yes! & \href{https://huggingface.co/datasets/walledai/AdvBench}{Link} \\
HarmBench \cite{mazeika2024harmbench} & Harmful QA & Harmful score & No& \href{https://www.harmbench.org/}{Link} 
\\ \midrule
AlpacaEval \cite{alpaca_eval} & General QA & Fine-tune accuracy & No &  \href{https://github.com/tatsu-lab/alpaca_eval}{Link}\\ 
MT-Bench\cite{zheng2024judging} & General QA & Fine-tune accuracy & No & \href{https://github.com/tatsu-lab/alpaca_eval}{Link} \\
TruthfulQA\cite{lin2021truthfulqa} & General QA & Fine-tune accuracy & No &  \href{https://huggingface.co/datasets/truthfulqa/truthful_qa}{Link} \\ 
ARC\cite{allenai:arc} &  Reasoning & Fine-tune accuracy & Yes! & \href{https://huggingface.co/datasets/allenai/ai2_arc}{Link}\\ 
GSM8K \cite{cobbe2021training} & Math & Finetune accuracy & Yes! &  \href{https://huggingface.co/datasets/openai/gsm8k}{Link} \\
GLUE (SST2) \cite{wang2018glue} & Classification & Finetune accuracy & Yes! &  \href{https://huggingface.co/datasets/nyu-mll/glue}{Link} \\ 
MMLU \cite{hendrycks2020measuring}  & General QA & Finetune accuracy  
& Yes! &  \href{https://github.com/hendrycks/test}{Link} \\
\bottomrule
\end{tabular}
}
\vspace{-0.4cm}
\end{table}

\vspace{-0.2cm}
\subsection{Models}
 State-of-the-art open-sourced LLM models are used as base model for evaluation. Popular choices include Llama2-7B \cite{touvron2023llama}, Llama3-8B\cite{meta2024introducing}, Qwen2 \cite{yang2024qwen2}, Gemma2\cite{team2024gemma}, OPT \cite{zhang2022opt}, Vicuna\cite{vicuna2023}, etc. Of note, usually there are two types of LLMs available, i.e.,  the aligned and the non-aligned version. For example, for Llama2, Llama2-7B is the pretrained model that is not safety aligned in advance, while Llama2-7B-chat is safety aligned by Meta in advance. Existing research (e.g., \cite{hsu2024safe, lyu2024keeping, qi2023fine, rosati2024representation, tamirisa2024tamper}) primarily use the aligned version (e.g., Llama2-7B-chat) as base model for evaluation, while a line of work by Huang et al. \cite{huang2024vaccine, huang2024lazy, huang2024antidote, huang2024booster} use unaligned version instead. We refer to the readers to Q4 in Appendix \ref{qoi} for how to choose the models.

\vspace{-0.1cm}
\section{Challenge and future direction}
\label{future}

\subsection{Attack side}

On the attack/red-teaming side, several research efforts can be made to advance the field. In the fine-tuning-as-a-service threat model, the attackers only have control over the fine-tuning data. Future research should be focused on constructing fine-tuning data to launch \textbf{stealthier} or \textbf{stronger} attacks.

\begin{itemize}[leftmargin=*]
\vspace{-0.2cm}
\item \textbf{Stealthier attack via steganography}. As shown in Table \ref{moderation mdoel}, most of the harmful data mixed in the user fine-tuning data might be filtered out by a moderation model. Factually, OpenAI seems to be adopting a moderation model to filter out the harmful data for their fine-tuning API. Therefore, to improve the attack's success rate, one may design a more stealthy attack method that might circumvent the moderation filtration. Steganography-based solution CMF \cite{halawi2024covert} is an attempt to reach this goal.  The methods basically construct encoded fine-tuning data to teach the model to answer the harmful question in an encoded way.  However,  in testing time, these encoding methods require that the question be encoded, and the harmful answer returned by the LLM is also encoded. This restriction might result in the loss of some generality of the attack (cannot be applied as evidence in the malicious case to sue the service provider for policy violation). Another technical challenge of this category of method is that the steganography fine-tuning data is still easy to detect because of its high perplexity. Subsequent study \cite{davies2025fundamental} aims to construct the encoding scheme with low perplexity to fill up this gap, but the proposed solution Flower is restricted in answering multiple choice questions (MCQ). Given the mentioned research gap of this line of study,  future research should be devoted to design an encoding scheme with lower perplexity, and can be extended to broader application scenarios. 

\item \textbf{Stealthier attack via guardrail jailbreak}. To evade guardrail detection, another attempt \cite{huang2025virus} aims to modify the harmful data by token-wise replacement, such that the optimizable harmful data can i) jailbreak the guardrail model, ii) effectively break down the guardrail. Different from \cite{halawi2024covert}, this method does not require testing time encoding/decoding. However, it is shown by their Table 9 that the Virus attack exhibits weak transferability in the blackbox setting, e.g., the Virus's data optimized on Granite Guardian3.1 cannot transfer well to Llama Guard.  Moreover, Virus only considers guardrail moderation in the fine-tuning data, but do not consider moderation over the model's input and output in the inference time.  Future research can be made to fill the mentioned research gap of i) relatively weak transferability, and ii) low generalization in the scenario of multi-phase guardrail moderation. 

\item \textbf{Stronger attack via better benign data sampling and construction}. Because the attacker might have limited harmful data samples to be able to pass through the moderation filtration, and finally used for fine-tuning, but \emph{the benign data can always pass through the detection}. For the attacker side, a stronger benign data sampling and data construction scheme might be useful to consider, in order to improve the attack performance.  \cite{he2024s} is an initial attempt for curating benign data with stronger attack performance. Following this line of study, \cite{guan2025benign} further builds an assumption-less data curation solution via self-influence function. On the other hand, \cite{kazdan2025no} designs a better benign data construction scheme to hide the affirmative answer to be a few tokens deeper, such that it can circumvent the safety alignment that is only a few tokens deep, as suggested by \cite{zhao2025weaktostrongjailbreakinglargelanguage,qi2024safety}. However, as shown in Table 1 of \cite{guan2025benign}, the attack performance of fine-tuning attack with benign samples still cannot outperform harmful data attack (under the case that no guardrail moderation is in place). Future research should be devoted in how to construct stronger benign data by better attack data sampling and construction scheme.   
\vspace{-0.3cm}
\end{itemize}

\vspace{-0.1cm}
\subsection{Defense side}
On the defense side, the following efforts might be worth exploring:
\begin{itemize}[leftmargin=*]
\vspace{-0.1cm}
   \item  \textbf{Data filtering/moderation-based defense}.
Existing defense literature focuses on improving the training method. However, it is also interesting to consider launching the attack right over the attack surface, i.e., to better filter out the harmful data from the user fine-tuning dataset when they are uploaded. One may look at the embedding space, and apply SOTA outlier detection method to better identify those harmful data among the benign user fine-tuning data. Recent research  SAFT \cite{choi2024safety} is a good attempt, but it is interesting to study how newer OOD methods perform.
 
\item \textbf{Prompt engineering}. While pioneer studies, e.g., PTST \cite{lyu2024keeping} and BEA \cite{wang2024mitigating} has explored the use of system prompt for defense design. It might be interesting to further extend on how to design alignment/fine-tuning/deployment system prompt such that the model will not learn from the harmful user prompt but effectively learn the fine-tuning knowledge from the benign user prompt.  
    \item  \textbf{Preference learning}. Most of the existing defenses focus on improving upon SFT. However, how to design a defense for RLHF based on preference data (a pair of good/bad answers for the same question) is still under-explored. Future defense design can explore how to better exploit the data structure of preference data to design stronger defenses.   

    \item  \textbf{Parameters-wise partial tuning/pruning. } Since the initial study from \cite{wei2024assessing} pointing out the observation that only a subset of parameters are responsible for safety function, there have been a surge of defense ideas developed based on this insight. For example, the alignment stage solution  RSN-Tune \cite{zhao2025identifying} identifies and only tunes (aligns) those safety parameters that do not overlap with foundation parameters. Another alignment stage solution T-Vaccine  \cite{liu2024targeted} argues that it is only necessary to add perturbation to some particular layers of parameters to strenghten the model's robustness. Fine-tuning stage solution ML-LR \cite{du2024towards} identifies the a robust subset of parameters and assigns a smaller learning rate to other parameters. Post-fine-tuning stage solution Antidote \cite{huang2024antidote} identifies and removes the harmful parameters after the fine-tuning. These research shares a strong but mysterious  connection, and there should be one empirical research to demystify the hidden principle behind those research, and proposes better solutions based on the obtained insight.   
      \item  \textbf{Memory/computation-efficient defense}. Most existing defenses increase the memory/computation overhead compared to SFT (without defense). A future research direction may be to utilize existing model acceleration techniques, e.g., pruning/sparsification \cite{frankle2018lottery,huang2022achieving,ilhan2024resource,tekin2024robust}, quantization \cite{dettmers2024qlora,li2023loftq} or factorization \cite{hu2021lora,huang2023fusion,li2023losparse} for a safety-efficiency co-design. 
   \item  \textbf{Defense by integrating multiple training stages}. Existing defenses against harmful fine-tuning typically focus on a single stage, such as the alignment stage. However, it is worth exploring the potential benefits of jointly optimizing across the alignment, fine-tuning, and post-fine-tuning stages to improve defense strategies. While this approach may introduce additional complexity, a co-design method could lead to more robust defense performance. Multi-stage training can also draw inspiration from successful defense techniques in multi-task learning and meta-learning~\cite{das2022skelevision, mao2020multitask}.
   \vspace{-0.1cm}
\end{itemize}

\vspace{-0.2cm}
\subsection{Mechanism interpretability study}
Interpretability study/explainable research could contribute to a better understanding of the problem, and eventually help the design of defense. We in the following summarize a few phenomena that were discovered by the existing work and discuss the challenges towards better understanding them.  
\begin{itemize}[leftmargin=*]
\item \textbf{Embedding drift}. It is first introduced in Vaccine \cite{huang2024vaccine} that fine-tuning on explicit harmful samples might cause embedding drift of the alignment data. It is still unknown whether this phenomenon is universal to all harmful fine-tuning attacks. In other words, the answer of whether all the harmful fine-tuning attack leads to embedding drift is still not definitive. 
\item \textbf{Layerwise/parameter-wise safety.} Safe LoRA \cite{hsu2024safe} highlights that certain layers of a model play a greater role in maintaining safety, while others are less susceptible and seem more critical for downstream tasks. Similarly, \cite{leong2024no} emphasizes that different layers serve distinct functions when exposed to various types of attacks. In vision domain, \cite{ilhan2024adaptive} also emphasize that different layers  serve different functions. In terms of safety functionality, Peng et al. \cite{peng2023robust} have established robust architectural design principles for adversarially robust CNNs. Drawing from insights in the vision domain, we argue that a layer-wise/parameter-wise mechanism analysis is essential to fully understand the role of different LLM layers in ensuring safety and to evaluate whether this principle holds universally. A more accurate way to identify the safety layer may be a by-product of analysis and may contribute to future defense design.

\item \textbf{Gradient ascent.} Several studies, e.g., RepNoise\cite{rosati2024representation}, TAR \cite{tamirisa2024tamper}, Booster\cite{huang2024booster} utilizes a gradient ascend term to unlearn the harmful knowledge (i.e., to maximize the loss over harmful data). However, from the authors' experience, a bad configuration of the gradient ascend term might negatively result in the model forgetting the normal knowledge as well. For example, the model may constantly output a single word for any given prompt after unlearning under bad configuration (e.g., hyper-parameters). It is worth studying in which case it will result in a negative effect and what is a better way to configure the gradient ascend term. 

\item \textbf{Safety basin}. It is discovered in \cite{peng2024navigating} that the aligned model exhibits a safety basin, and harmful fine-tuning is in essence dragging the model weights out of this basin. It is interesting to study how the models produced by different defenses are going to shape the safety basin, and how perturbation yielded by different attacks can drag the model out of the safety basin. 
\end{itemize}

A comprehensive benchmark could also be of great value to contribute and may advance the development of attack/defense design. 
\begin{itemize}[leftmargin=*]
\item 
\textbf{Attack/Defense Benchmark.} Existing research papers on attack/defense towards harmful fine-tuning generally have different experimental methodologies. It is imperative to create a standard benchmark for modeling the attack/defense performance of the existing/future solutions. The benchmark may provide options for SFT and other RLHF methods (e.g., PPO, DPO) for alignment and user fine-tuning. \cite{rosati2024defending,hossain2026tamperbenchsystematicallystresstestingllm,wei2025best} are initial attempts but still can  be further improved for their comprehensiveness. 
\end{itemize}

\vspace{-0.1cm}
\section{Conclusion}
We have systematically introduced the recent harmful fine-tuning attack against safety-aligned models in the context of fine-tuning-as-a-service business platform. We also provide the characterization of existing defense methods in the literature, and introduce the current experiment methodology for risk assessment and mitigation effectiveness. We conclude the survey with discussions on some future research directions and the challenges ahead.
We conjecture that this survey will help LLM researchers and engineers to gain in-depth understanding of harmful fine-tuning attacks and defenses, and will also serve as a timely guidance for future research endeavors towards fortifying the safety aligned large language and vision models in anticipation of small harmful fine-tuning data samples. 


\vspace{8pt}
\noindent {\bf Acknowledgement.} 
This research is partially sponsored by the NSF CISE grants 2302720, 2312758, 2038029,  an IBM faculty award, and a grant from CISCO Edge AI program. The first author is 
sponsored by Google PhD fellowship 2025, and he 
would like to thank Domenic Rosati, Shengyun Peng, Xiangyu Qi, Boyi Wei, Mantas Mazeika, Rui Ye,  Apurv Verma, Hyeongkyu Choi, Stephen Casper,  and Han Shen  for the  discussion when working on this survey project.


\bibliographystyle{ACM-Reference-Format}
\bibliography{sample-base}

\medskip
\newpage
\appendix
\begin{center}
    \LARGE\bfseries E-pub Only Appendix 
\end{center}
\section{Question of Interest}
\label{qoi}
We below aim to answer questions that might be of interest to the readers. The answers only reflect the authors' personal opinions and may not be the norms for this research problem.

\textbf{Q1: How to evaluate the harmful score of the model after fine-tuning?}

As surveyed in Section \ref{exp setup}, model-based evaluation is the mainstream for evaluating how harmful the model is. Among model-based evaluations, there are two categories of evaluation methods. The first method is to evaluate with GPT4's API, i.e., to prompt GPT4 to rate the score to the model's output. The second method is to use an open-source moderation model to flag the harmful answers and calculate the percentage of the answers to be flagged as harmful. 

Compared to the second method, the GPT4 evaluation method has the following advantages. 
\begin{itemize}[leftmargin=*]
    \item \textbf{Well-accepted by the research community}. Safety alignment research mostly uses GPT4 for evaluating harmful scores, and the community has accepted this as the mainstream evaluation method. Therefore, it has less chance to be questioned in the peer-review process. 
    \item \textbf{High accuracy}. The GPT4 evaluation can accurately reflect the human evaluation.
\end{itemize}

However, the GPT4 method has an important downside.

\begin{itemize}[leftmargin=*]
    \item \textbf{Financially Costly}. Calling GPT4 API to evaluate the model's output may be financially expensive, and may not be affordable for many research groups around the globe.  
\end{itemize}

The GPT4 evaluation method is used in \cite{qi2023fine}, while the second evaluation method is used in \cite{huang2024vaccine}.


\textbf{Q2: What fine-tuning task should I use and how to evaluate the fine-tuning task performance?}

Existing research papers use very different fine-tuning task settings. A common dataset to use include MT-Bench, AlpacaEval, and GSM8K.


\textbf{Q3: Should I use RLHF (e.g., PPO, DPO), SFT(Full) or SFT (LoRA) for alignment/fine-tuning?}

The quick answer is that the priority is RLHF> SFT (Full) > SFT (LoRA). RLHF has shown its superiority over SFT. Research directly based on RLHF will be more practical and of interest to the broad community. However, the GPU resource requirement of RLHF is significantly larger than SFT (Full) and  SFT (Full) is significantly larger than SFT (LoRA). Also, the training time of SFT (LoRA) will be sufficiently smaller than the two other methods, due to its efficient training nature. 

\textbf{Q4: Should I use the aligned version of LLMs (e.g., Llama2-7B-chat) or the non-aligned version (e.g., Llama2-7B) for the experiment?}

Using an aligned model (e.g., Llama2-7B-chat), rather than aligning a Llama2-7B yourself, would be simpler for developing a fine-tuning stage and post-fine-tuning stage solution. However, because we as researchers are agnostic to the alignment process, it may contain the risk that the defense is effective only because some unknown procedures are taken in the alignment process (i.e., when Llama2-7B-chat is produced). For an alignment stage solution, it may be more reasonable to use the unaligned model as the base model, because that means the defense can be integrated into the model's safety alignment process (which needs to be done by all the service providers, e.g., Meta, OpenAI). 


\textbf{Q5: What kind of assumptions of available datasets should we make when we design a defense?}
As illustrated in \ref{exp setup}, the assumptions of the dataset that existing papers have made include: i) safety alignment dataset (harmful question-safe answer pair). ii) in-distribution harmful dataset (harmful question-harmful answer pair, but they are different instances used in the harmful-finetuning process). iii) Proxy dataset (Safe general question-safe answer pair). 


\textbf{Q6: Which name of metrics should we use for paper writing?}

As the research of harmful fine-tuning attacks is still in the early stage, different papers use different names of metrics for measuring the harmfulness of the model as well as the fine-tuning tasks' accuracy. We advocate using \textbf{harmful score} and \textbf{fine-tune accuracy}, as people seem to be able to immediately get the meaning of the two metrics by their name, and they are short such that can be easier to display in a table. 

\textbf{Q7: Why do we consider the fine-tuning-as-a-service scenario, instead of the scenario that the user herself fine-tunes the model?}

Several studies, e.g., \cite{tamirisa2024tamper, rosati2024representation} consider improving the aligned model's robustness, such that users cannot break the alignment even though they have full control of the model weights, the fine-tuning process, as well as the deployment process. 

However, there are two critical issues with this setting. 

Particularly, what is the motivation that attackers want to attack their own model, and why the service provider needs to worry about that? The answer to the first question might be that users unintentionally include harmful data into the fine-tuning data. However, in this case, why do the service providers need to worry about the alignment-broken of the fine-tuned model? The models are fine-tuned by the users using users' data and are deployed by the users themselves. The service provider does not clearly have responsibility for the output of the fine-tuned model. 

Another issue is that the defense seems to be very hard in this setting.  In an extreme example, the users already have an alignment-broken model, and they claim that this alignment-broken model is fine-tuned from the service provider's release model. Then any defenses can fail in this case. Indeed, the defense is hard even in normal settings because the users have full control over the fine-tuning process and they might adopt larger optimization steps to subvert the aligned model.   


\section{Related survey}
\label{related work}
We group the existing surveys on LLM safety into three broad categories and hightlight our unique contribution to the community. 

\textbf{The first category focuses on a general discussion of security, privacy, and trust of LLMs from both vulnerabilities and mitigation methods.} 
Zhao et al. \cite{zhao2024survey} systematically review the backdoor attack/defense for LLMs and classify classify backdoor attacks into three categories: full-parameter fine-tuning, parameter-efficient fine-tuning, and no fine-tuning. 
Yao et al. \cite{yao2024survey} studies how LLMs positively impact security and privacy, and the potential risks and threats associated with their use, as well as inherent vulnerabilities within LLMs. 
Chua et al. \cite{chua2024ai} provides an up-to-date survey of recent trends in AI safety research. 
\cite{liu2023trustworthy} covers seven major categories of LLM trustworthiness: reliability, safety, fairness, resistance to misuse, explainability and reasoning, adherence to social norms, and robustness. 
He et al. \cite{he2024emerged} provide a comprehensive overview of the newly emerged privacy and security issues faced by LLM agents. 
Dong et al. \cite{dong2024safeguarding} provides a systematic literature review on the current status of LLM safeguard, and point out several challenges that cannot be easily addressed by the existing LLM safeguard. 
Das et al. \cite{das2024security} provides a thorough review of the security and privacy challenges of LLMs for both training data and users as well as the application-based risks in various domains. 
Shayegani et al. \cite{shayegani2023survey}  provide an overview of large language models, describe their safety alignment, and categorize existing research based on various learning structures: textual-only attacks, multi-modal attacks, and additional attack methods. Reuel et al. \cite{reuel2024open} explained what technical AI governance is and why it is important with a taxonomy of open problems.

\textbf{The second category focuses on analyzing risks of LLMs and mitigation approaches through evaluation platforms.} For example, 
Huang et al. \cite{huang2024survey2} reviewed the known vulnerability of LLMs and classified them  into inherent issues, attacks, and unintended bugs, and showed how Verification and Validation (V\&V) techniques can be integrated to provide analysis of the safety of LLMs. Dong et al. \cite{dong2024attacks} provides a comprehensive overview of recent studies on LLM conversation safety in terms of not only attacks and defenses but also evaluation methods. 
Rottger et al. \cite{rottger2024safetyprompts} conduct a first systematic review of open datasets for evaluating and improving LLM safety. Ayyamperumal et al. \cite{ayyamperumal2024current} explored the risks associated with deploying LLMs and evaluated current approaches to developing guardrails and model alignment techniques. 
Cui et al. \cite{cui2024recent} reviewed current research on LLM vulnerabilities and threats, and evaluated the effectiveness of contemporary defense mechanism.

\textbf{The third category mainly dedicates to the attacks/redteaming towards LLMs.} Concretely, numerous specialized reviews are provided about jail-break attacks against LLMs. Representative ones include Xu et al. \cite{xu2024comprehensive}, Chowdhury et al. \cite{chowdhury2024breaking}, Chu et al. \cite{chu2024comprehensive}, Jin et al. \cite{jin2024jailbreakzoo}, Yi et al. \cite{yi2024jailbreak}, Peng et al. \cite{peng2024jailbreaking}, Liu et al. \cite{liu2024exploring}. 
Shen et al. \cite{shen2023large} provides an extensive review of safety alignment methodologies for LLMs. 
Verma et al. \cite{verma2024operationalizing} provide a systematic review of red-teaming attacks on LLMs, including the jailbreak attack, direct attack, infusion attack, inference attack, as well as one type of harmful fine-tuning attack (alignment erasure in their paper).

From the above discussion, existing survey papers exhibit the following research gap:
\begin{itemize}[leftmargin=*]
   \vspace{-0.1cm}
    \item Existing surveys mostly focus on a \textbf{general} discussion on different kinds of risks (threat models) but lack in-depth analysis of the existing threat model/relevant literature. 
    \item Due to the diversified threat models they are considering, existing survey papers only emphasize one aspect (e.g., attack, evaluation) on the safety of LLM, but cannot derive a focused explanation of all the aspects (i.e., attacks/defenses/evaluations) of one particular threat model.  
    \vspace{-0.1cm}
\end{itemize}

To fill the research gap, our survey specifically focuses on a systematic revision of harmful fine-tuning -- a newly emerging threat of LLM harmful fine-tuning. To the best of our knowledge, this paper is the first attempt (along with a concurrent study \cite{casper2026open}) to present a systematic review of harmful fine-tuning.  With a specific focus on \textbf{one type of threat model}, we are able to provide a dedicated illustration of the attack setting, a comprehensive coverage of relevant existing attacks/defenses, as well as a detailed introduction of the evaluation methodology for harmful fine-tuning. In addition, our paper proposes a unique taxonomy of existing defenses against harmful fine-tuning (i.e., alignment stage, fine-tuning stage, post-fine-tuning stage) based on the timing at which the defenses take place, which is not available beforehand. With these unique features, we trust that our survey provides a different reading experience compared to surveys that cover the general topics of LLM safety. We acknowledge that this survey may not be comprehensive enough to cover all threat models for LLMs (interested readers can refer to the related survey papers above).

\newpage

\section{More Statistical Results}
\label{more results}
\begin{figure}[!h]
    \centering
    \vspace{-0.2cm}
    \includegraphics[width=1\linewidth]{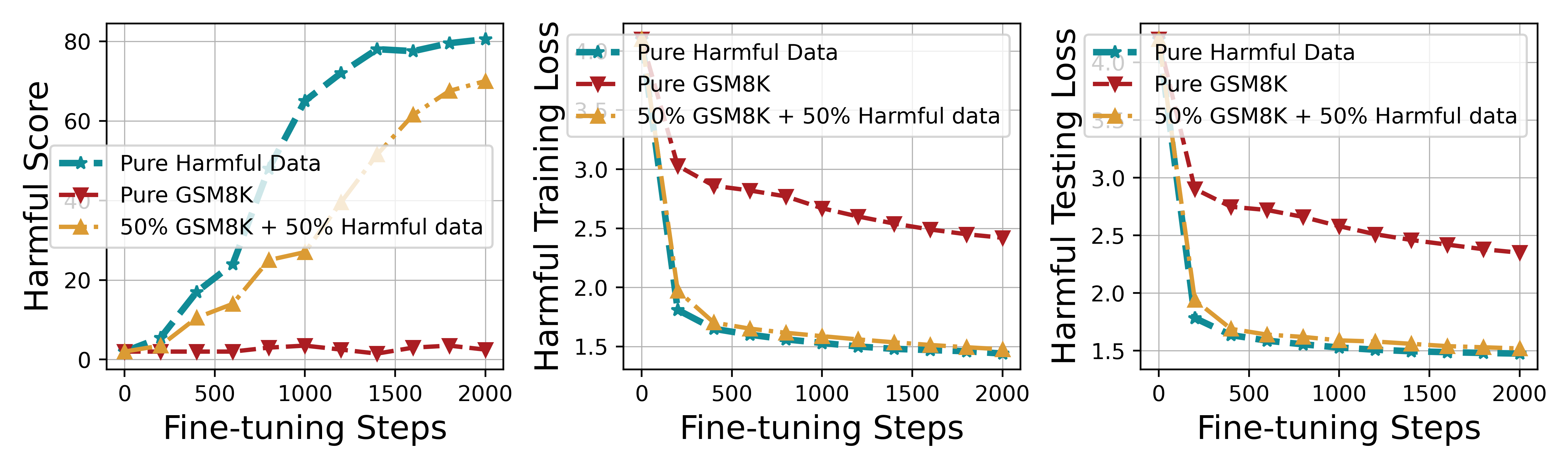}
    \vspace{-0.8cm}
    \caption{Model statistics (Left: harmful score, Middle: harmful training loss, Right: harmful testing loss) after fine-tuning  on pure GSM8k/harmful data as well as 50\% mixture data for different steps. Specially, harmful score measures how harmful the model is (the smaller the better), harmful training loss refers to the loss over the harmful data used in fine-tuning, while harmful testing loss refers to that over the testing harmful data that the model never sees in fine-tuning stage.   }
    \label{motivation gsm8k}
\end{figure}

In Section \ref{statistical findings}, we visualize the harmful score and training loss with respect to training steps when the model is fine-tuned on SST2 dataset, which is a typical classification dataset. Next, we do an experiment on a question-answering (QA) dataset GSM8K to show the generalization of the findings. We show the results in Figure \ref{motivation gsm8k}, and we derive the following findings, which are mostly consistent with those derived from SST2: i) Harmful score is increasing with more fine-tuning steps when fine-tuning on pure harmful/partially harmful data, but is not significantly increasing when fine-tuning on pure GSM8K dataset. ii) The harmful training loss and harmful testing loss 
are decreasing with more fine-tuning steps for all sets of fine-tuning data, though the extent of decrease with GSM8K is slighter than that of the pure harmful/partially harmful data.  This result is not consistent with the result when fine-tuning on SST2 (Figure \ref{motivation}), which shows that fine-tuning on benign dataset SST2 increases the harmful training/testing loss. The reason for such discrepancy is probably that GSM8K more resembles a harmful dataset (which is a QA dataset as well), making the gradient descent on GSM8K also reduce the harmful loss. On the contrary, the SST2 is a classification dataset, which may not share much similarity with the harmful dataset, resulting in an increase of harmful loss.

\section{Other Safety Unalignment Attack Scenarios}
\label{attack settings}
Except harmful fine-tuning attack for fine-tuning-as-a-service, there might be other scenarios for safety unalignment attack. An important attack surface is over the open-source model. 

\begin{table}[!h]
\centering
\caption{Overview of attackers' and defender's capability for different attack scenario. SA and FT respectively means the permission of safety alignment and the fine-tuning process. \ding{108}  means partial control.}
\resizebox{1\linewidth}{!}{
\begin{tabular}{c|cccc|cccc}
\toprule
Scenarios    & \multicolumn{4}{c}{Attacker's capability} & \multicolumn{4}{c}{Defender's capability} \\ 
                                     & SA      & FT     & Deploy     & Infer     & SA      & FT     & Deploy     & Infer     \\ \midrule
Fine-tuning-as-a-service (user as an attacker)      &     \ding{55}    &       \ding{108}   &  \ding{55}            &   \ding{51}          &      \ding{51}    &   \ding{108}      &          \ding{51}     &   \ding{55}           \\  \midrule
Jail-break (user as an attacker)      &     \ding{55}    &       \ding{55}   &  \ding{55}            &   \ding{51}          &      \ding{51}    &   \ding{51}      &          \ding{51}     &   \ding{55}           \\  \midrule
Opensource (user as an attacker)      &     \ding{55}    &       \ding{51}   &  \ding{51}            &   \ding{51}          &      \ding{51}    &   \ding{55}      &          \ding{55}     &   \ding{55}           \\  \midrule
Opensource (publisher as an attacker) &     \ding{51}       &       \ding{51}    &        \ding{55}        &    \ding{55}           &       \ding{55}      &    \ding{51}        &      \ding{51}          &     \ding{51}         \\   \bottomrule            
LLM service with backdoor (publisher as an attacker)      &     \ding{51}    &       \ding{51}   &  \ding{51}            &   \ding{55}          &      \ding{55}    &   \ding{55}      &          \ding{55}     &   \ding{51}           \\  \midrule

\end{tabular}
}
\end{table}

\subsection{Risk of Open-source model: model users as an attacker}
\textbf{Scenario description}. Existing LLM open-source model's publishers (e.g., Meta, Microsoft) release safety-aligned versions of LLMs (e.g., Llama2-7B-chat) on open-source platforms (e.g., Huggingface). However, the safety alignment of this aligned LLMs can be compromised if the users fine-tune the model on harmful or seemingly benign data. This attack scenario is adopted by RepNoise \cite{rosati2024representation} and TAR \cite{tamirisa2024tamper}. 

\textbf{Attack motivation}. The attackers (users) want to unalign the model because they want to accuse/blame the model publisher for the weak alignment. Indeed, it is possible that fine-tuning demonstration data only serves to elicit the harmful behaviors of the models. The root of the evil behaviors of the LLM is that the model publisher does not filter harmful data for pre-training, and they feed this harmful data to the LLM and grow this evil intelligent "creature". The model publisher has partial liability for the harmful behavior of the model.    

\textbf{Attacker capability}. 
Attackers in this scenario are the users. The users have full knowledge of the open source model's weights, and can arbitrarily manipulate the model. For example, they can do a harmful fine-tuning attack \cite{yang2023shadow} to elicit the model, or they can also do a weights replacement attack \cite{li2024badedit}. They can also manipulate the model inference process, e.g., applying TA\cite{wang2020attack}.  

\textbf{Defender capability}. Defenders can only manipulate the safety alignment process to make it more robust for attack. It has no control over the fine-tuning/inference process. The authors are pessimistic towards defense in this scenario, given the evil nature of the pre-training LLMs--they are trained with a massive amount of data on the Internet, and maybe a large portion of them could come from nasty websites.

\subsection{Risk of open-source model: model publisher as an attacker}
\textbf{Scenario description}. On the other hand, attackers might also be the open-source model publisher. A justified attack could be a backdoor unalignment attack. Consider that the model publisher publishes a backdoor model that has a backdoor, which can be implanted with fine-tuning attack \cite{shi2023badgpt}. In testing time, the safety alignment of the model will be subverted whenever a backdoor trigger is presented, but the users have no awareness of the backdoor because they are unaware of the backdoor trigger. The victim users might deploy the models in their servers, and subsequently be attacked by the model publisher.

\textbf{Attack motivation}. The attackers (model publishers) want to backdoor and unalign the model because they want to accuse/blame the users for providing harmful outputs. Because the models are deployed by the model users, and the harmful answers are transmitted from the users' API. Users have liability for the model's output.  According to the statement in the previous scenario, it is possible that the attacker (i.e., model publisher) have also partial liability for the model's output. However, what if the model publisher is someone anonymous (or someone-no-one-care), but the model user is Microsoft? Will the attacker care about his liability for attacking Microsoft?  

\textbf{Attacker capability}. 
The attacker has full control over the alignment and backdoor fine-tuning process. Once the open-source model is released, the attacker cannot perform any operation on the model.  

\textbf{Defender capability}. The defender has full control over the model deployment/inference process but has no control over the alignment. They can also do further fine-tuning on the model to erase the backdoor.  The defense of attack in this scenario is more realistic than the previous scenario. because the defender still has control on what model she is going to deploy as well as its inference process.

\subsection{Risk of LLM service with backdoor: Publisher as an attacker}
\textbf{Scenario description}.  Due to the computing resources needed for LLM model deployment and privacy/security concerns of the model publisher, there is a growing trend that the model publisher chooses to deploy the LLMs on the cloud (e.g., Amazon cloud, Huggingface), and only inference API is exposed to model users.  Model users might use this inference API to develop applications and provide service to application users, e.g., ~\cite{gptlens,hu2024survey,hu2023bert4eth,hu2024zipzap}. However, this paradigm poses a serious safety threat to the users, given the black box nature of the service. For example, the LLM being deployed might contain a backdoor, in that only when the backdoor trigger is presented, the safety alignment of the model will break down. Because the model users are unaware of the backdoor trigger, they cannot be aware of risk even if an evaluation is done prior on the API. Once the application is released, the model publisher might use the trigger to attack the model users.

\textbf{Attack motivation.} The attackers (model publisher) want to backdoor and unaligned the model because they can accuse the users of providing harmful outputs. Because the model users release an application to the application users, the model users should have liability for the content presented in their application. In this scenario, it is true that the model publisher should also have liability. But again, the model publisher can be anonymous or may not know/care the legal consequence of her act.      

\textbf{Attacker capability}. 
The attacker has full control over the alignment, backdoor fine-tuning process, and deployment process. However, once the model is deployed, they lose control over the model.  

\textbf{Defender capability}. The defender does not know the model weights, and neither they have access to the alignment, backdoor fine-tuning process, and deployment process. The defender can only call the inference API for defense.

\section{Relevant topics for LLM safety}
We in this section provides more background on relevant papers on the broad topic of LLM safety. 

\subsection{Safety alignment on LLM}
\label{safety alignment}
 Safety alignment focuses on training a large language model (LLM) to generate outputs that are both helpful and harmless, in line with human preferences. A key component of this process is the human-aligned supervised dataset, which plays a crucial role in ensuring safety alignment. The challenge lies in how to effectively leverage this alignment dataset. Techniques based on Reinforcement Learning from Human Feedback (RLHF) \citep{ouyang2022training, griffith2013policy, dai2023safe, bai2022training, wu2023pairwise, dong2023raft, rafailov2023direct, yuan2023rrhf, song2023preference} often use pairs of preference data to guide the alignment process. A classic example is the original Proximal Policy Optimization (PPO) design, which uses supervised fine-tuning (SFT) to train a reward model based on the preference dataset. The reward model then provides a supervised signal to the pre-trained LLM during the subsequent alignment phase. Other alignment methods include Chain of Hindsight \citep{liu2023chain}, which employs pairs of good and bad answers for Supervised Fine-Tuning (SFT), as well as Stable Alignment \citep{liu2023training} and Selfee \citep{ye2023selfee}, all of which enhance alignment data by using prediction and re-evaluation. A more recent work \citep{zou2024improving} introduces circuit breakers, which train a LoRA adaptor, such that the hidden representation of harmful input is mapped to an orthogonal direction of the original representation.  Tekin et al. \citep{tekin2024h} proposes $H^3$Fusion, which utilizes a Mixture of Experts (MoE) method to combine three types of safety aligned models (Helpfulness, Harmlessness, and Honesty) through a learning based fusion model, further boosting the alignment performance on creating helpful, harmless and honest response.  

\subsection{Other attack/defenses towards LLM}
\label{other attacks}
 While this paper primarily focuses on the recent harmful fine-tuning attacks, we in this section slightly extend the discussion to other attack scenarios.

\textbf{Jail-break Attack/Defense}. It is discovered in \cite{wei2024jailbroken,zou2023universal, shen2023anything} that it is possible the circumvent the safety alignment of a model by manipulating the user prompt. Specifically, \cite{wei2024jailbroken,shen2023anything} utilize carefully crafted manual prompt to elicit the harmful answer,  \cite{zou2023universal} propose GCG attack to automatically produce the adversarial suffix for each prompt for jail-breaking. However, the GCG attack can be easily circumvented by a naive perplexity defense, while the manual prompts in \cite{wei2024jailbroken,shen2023anything} may not be effective. Therefore, AutoDan \cite{liu2023autodan} utilizes a Genetic algorithm to produce a suffix that is both readable and achieves good defense performance.  A similar idea to mutate and select a better readable prompt is also utilized in Gptfuzzer \cite{yu2023gptfuzzer}. To achieve the same goal, Pair \cite{chao2023jailbreaking} uses an attacker LLM to automatically generate a readable jailbreak prompt for the to-be-attacked LLM, and TAP \cite{mehrotra2023tree} further introduces an evaluator (another LLM) assesses generated prompts and prunes the ones unlikely to result in jailbreaks before sending them to the to-be-attacked LLM.

On the defense side, smoothLLM \cite{robey2023smoothllm} generates multiple prompts by randomly perturbing the input prompt (potentially contains jailbreak suffix), and uses majority voting to decide the final answer to the prompt, and a concurrent study RALLM \cite{cao2023defending} also utilize a similar random perturbation techniques to decide the final answer.   Bergeron \cite{pisano2023bergeron} uses secondary LLM acting as a guardian to protect the target LLM. \cite{jain2023baseline} apply several baseline defenses originally proposed in the computer vision domain against jail-break attacks, e.g., adversarial training. LLM-Self-Defense~\cite{phute2023llm} is a simple approach to defend against jailbreak attacks by having an LLM screen the induced responses, which does not require
any fine-tuning, input preprocessing, or iterative output generation. \citep{candogan2024single} propose SPD, which achieve defense by classifying the jailbreak samples based on their output logits.   RPO \citep{zhou2024robust} achieves defenses by first simulating the attack to obtain jailbreak suffix, and then optimize a defensive suffix to counter the obtained jailbreak suffix. 

For more information, we refer to a comprehensive comparison study in \cite{chu2024comprehensive}, and a survey in \cite{yi2024jailbreak}.

\textbf{Backdoor Attack/Defense}.  Backdoor attack allows an attacker to manipulate the model's output whenever a pre-defined trigger is presented in its input.  Backdoor attack for deep learning model is proposed in \cite{gu2017badnets} and then is extended to multiple scenarios, e.g., federated learning \cite{bagdasaryan2020backdoor,huang2024lockdown}, diffusion model \cite{chou2023backdoor,pan2023trojan}, etc.

Standard backdoor attacks require a backdoor dataset and require the model to train on this dataset to launch an attack. The first backdoor attack for LLM is BadGPT \cite{shi2023badgpt}. To inject an attack into the model, BadGPT assumes that the attackers might present a backdoor dataset, and the following RLHF trained on this dataset might make the model start to exhibit backdoor behaviors. Subsequent research \cite{yan2024backdooring} proposes VPI to show that a specific keyword of a topic can used as a trigger for a backdoor attack. The evaluation is done with SFT with a more comprehensive evaluation and analysis. \cite{chen2024dark} proposes a better prompt trigger selection method to enhance the attack performance. \cite{huang2023composite} proposes CBA, which implants multiple trigger keys in different prompt components. Authors in \cite{hubinger2024sleeper} propose Sleeper with "Current year: 2024" as a backdoor trigger, and demonstrate that the backdoor, once implanted, cannot be erased by safety alignment performed later. 

Another type of backdoor attack for LLMs does not require backdoor data for fine-tuning/training. For example, \cite{xiang2024badchain} proposes a backdoor attack BadChain for Chain-of-Thought (COT) reasoning\cite{wei2022chain}, which does not require access to training data but injects the backdoor by prompting through the COT process.  \cite{wang2023backdoor} proposes TA, which implants backdoor trigger by adding malicious vectors into the activation layers of an LLM in inference time.  BadEdit\cite{li2024badedit} implants the backdoor by directly editing the model weights.  

On the defense side, Beear \cite{zeng2024beear} apply the embedding drift idea and the perturbation-aware training from Vaccine \cite{huang2024vaccine} into backdoor defense for LLMs. Another research targeted LAT is \cite{sheshadri2024targeted} built upon the framework of \cite{casper2024defending}. They assume that the backdoor objective of the attacker (e.g., compromise the safety alignment) is known. Their idea to defend is to search for the perturbation such that the harmful loss with this perturbation is minimized. Then they train the model to be resistant to such perturbation in order to unlearn  association with backdoor trigger and harmful knowledge after the backdoor has been implanted.   

For more information on backdoor attack/defense for LLMs, we refer to \cite{li2024backdoorllm}, which provides a comprehensive benchmark for backdoor attack for LLMs.  For a more dedicated survey of backdoor attacks/defenses, we refer to \cite{zhao2024survey}. 

\textbf{Data Extraction attack/defense}. It is demonstrated by Carlini et al. \citep{carlini2021extracting} that one can recover individual training examples from a large language models (GPT2) by presenting a large amount of query to the model. 
Lehman et al. \citep{lehman2021does} ran a battery of experiments, attempting to recover sensitive training data from a Bert model. Their results show that while simple methods are able to recover the data, the more sophisticated attack based on generating text might be able to recover the sensitive training data. Carlini et al. \citep{carlini2022quantifying} shows that the memorization of the training data is closely related to the capacity of the model, the number of times an example has been duplicated and the number of tokens of context to prompt the model. Kandpal et al. \citep{kandpal2022deduplicating} further shows that the success of data extraction attack is also closely related to the duplication of the web-scrapped training data when they are used to pre-train the model. Jagielski et al. \citep{jagielski2022measuring}  provides a way to measure forgetting of training data and its association with the data extraction attack. Jayaraman et al.  \citep{jayaraman2022active} proposes an active extraction attack and evaluate the solutions with Bert and GPT2. 
Subsequent  research by Nasr et al. \citep{nasr2023scalable} shows that the training data extraction attack can also be effective to a safety aligned model, e.g., Chat-GPT. They found that by instructing the model to repeat a word endlessly, the model will final "diverges" and starts to generate  the training data. Another research by Yu et al.  \cite{yu2023bag} provides several optimization tricks to ensure a more successful attack, e.g., how to select a sampling strategy for decoding, how to address the sampling temperature or imposing repetition penalty. Ozdayi et al. \cite{ozdayi2023controlling} propose a prompt-tuning attack which optimize a pre-fix that increase the data extraction rate. Ozdayi et al. further propose a gradient ascent-based defense solution for the defender to mitigate the proposed prompt-tuning attack. Patil et al. \citep{patil2023can} show that even the SOTA model editing method may not be a successful defense toward data extraction attack, as the attacks are sill able to recover "deleted" information from the editted model. We refer to \citep{verma2024operationalizing} and \citep{brown2022does} for good surveys of the existing data extraction attack/defense.

\textbf{Data poisoning attack. } Data poisoning attack \cite{steinhardt2017certified, tolpegin2020data} (or label flipping attack) has been extensively studied in both language and vision domains. In the vision domain, vanilla data poisoning attack assumes that the attacker mixes some poisoning data with wrong label into the training data, and the model trained on poisoning data will mis-classify the input, i.e., be poisoned. Data poisoning attacks share some similarity and also exhibit some fundamental differences with harmful fine-tuning attacks, as summarized below.

\begin{itemize}[leftmargin=*]
\item \textbf{Similarities}. Both attacks share a similar assumption: the malicious data is mixed with benign training data and \textit{they cannot be separated with ease}. Their goal of both attacks share some similarities; they both aim to poison the model such that the model would give a target output whenever the model recognizes some specific pattern in the input.   

\item \textbf{Differences}. i) \textbf{The attack goals of two attacks might be different.} For example, backdoor attack is a type of data poisoning attack that aims to trigger a specific malicious behavior of the model. The scope of malicious behavior triggered by backdoor attack could be very diverse, e.g., to degrade safety alignment \citep{yi2025probe,zeng2024beear}, bypass the COT thinking of a reasoning model \cite{zhu2025think}, etc.  On the other hand, the goal of  harmful fine-tuning attack is strictly tied to break down the safety alignment of the model.  ii) \textbf{The input pattern triggered the malicious behaviors of the model could be different.} For example, for a backdoor attack, the attack is triggered when the model encounters a predefined input pattern, i.e., a backdoor trigger. On the other hand, harmful fine-tuning attack aims to elicit harmful behavior (i.e., output harmful answers) when the model is given a "harmful prompt" as input.
\end{itemize}

\section{More discussion on survey methodology}

\subsection{Paper collection record}

We  attach the paper collection record, which record the date and the tools we use for collecting the papers.

\begin{table}[!h]
\caption{Paper collection record.}
\resizebox{1\linewidth}{!}{
\begin{tabular}{ccc}
\toprule
Title                  & Collected date & Collected capacity \\
\midrule
Vaccine: Perturbation-aware alignment for large language model aginst harmful fine-tuning                            & 9/20/2024      & Google Scholar     \\
Representation noising effectively prevents harmful fine-tuning on LLMs                                              & 9/20/2024      & Google Scholar     \\
Robustifying Safety-Aligned Large Language Models through Clean Data Curation                                        & 9/20/2024      & Google Scholar     \\
Tamper-Resistant Safeguards for Open-Weight LLMs                                                                     & 9/20/2024      & Google Scholar     \\
Booster: Tackling harmful fine-tuning for large language models via attenuating harmful perturbation                 & 9/20/2024      & Google Scholar     \\
Fine-tuning can cripple your foundation model; preserving features may be the solution                               & 9/20/2024      & Google Scholar     \\
Safety-Tuned LLaMAs: Lessons From Improving the Safety of Large Language Models that Follow Instructions             & 9/20/2024      & Google Scholar     \\
Safety fine-tuning at (almost) no cost: A baseline for vision large language models                                  & 9/20/2024      & Google Scholar     \\
Assessing the brittleness of safety alignment via pruning and low-rank modifications                                 & 9/20/2024      & Google Scholar     \\
Mitigating fine-tuning jailbreak attack with backdoor enhanced alignment                                             & 9/20/2024      & Google Scholar     \\
Keeping llms aligned after fine-tuning: The crucial role of prompt templates                                         & 9/20/2024      & Google Scholar     \\
Lazy safety alignment for large language models against harmful fine-tuning                                          & 9/20/2024      & Google Scholar     \\
Safety alignment should be made more than just a few tokens deep                                                     & 9/20/2024      & Google Scholar     \\
A safety realignment framework via subspace-oriented model fusion for large language models                          & 9/20/2024      & Google Scholar     \\
Safe lora: the silver lining of reducing safety risks when fine-tuning large language models                         & 9/20/2024      & Google Scholar     \\
Antidote: Post-fine-tuning safety alignment for large language models against harmful fine-tuning                    & 9/20/2024      & Google Scholar     \\
Shadow Alignment: The Ease of Subverting Safely-Aligned Language Models                                              & 9/20/2024      & Google Scholar     \\
Fine-tuning aligned language models compromises safety, even when users do not intend to!                            & 9/20/2024      & Google Scholar     \\
Removing RLHF Protections in GPT-4 via Fine-Tuning                                                                   & 9/20/2024      & Google Scholar     \\
Covert malicious finetuning: Challenges in safeguarding llm adaptation                                               & 9/20/2024      & Google Scholar     \\
Lora fine-tuning efficiently undoes safety training in llama 2-chat 70b                                              & 9/20/2024      & Google Scholar     \\
No two devils alike: Unveiling distinct mechanisms of fine-tuning attacks                                            & 9/20/2024      & Google Scholar     \\
Navigating the safety landscape: Measuring risks in finetuning large language models                                 & 9/20/2024      & Google Scholar     \\
Do as I do (Safely): Mitigating Task-Specific Fine-tuning Risks in Large Language Models                             & 10/06/2024     & Google Scholar     \\
SEAL: Safety-enhanced Aligned LLM Fine-tuning via Bilevel Data Selection                                             & 10/06/2024     & Google Scholar     \\
Identifying and Tuning Safety Neurons in Large Language Models                                                       & 10/06/2024     & Google Scholar     \\
Bi-Factorial Preference Optimization: Balancing Safety-Helpfulness in Language Models                                & 10/06/2024     & Google Scholar     \\
Safety Alignment Shouldn't Be Complicated                                                                            & 10/06/2024     & Google Scholar     \\
Locking Down the Finetuned LLMs Safety                                                                               & 10/06/2024     & Google Scholar     \\
Your Task May Vary: A Systematic Understanding of Alignment and Safety Degradation when Fine-tuning LLMs             & 10/06/2024     & Google Scholar     \\
Making Harmful Behaviors Unlearnable for Large Language Models                                                       & 10/06/2024     & Google Scholar     \\
Language Models are Homer Simpson! Safety Re-Alignment of Fine-tuned Language Models through Task Arithmetic         & 10/06/2024     & Google Scholar     \\
Defending against Reverse Preference Attacks is Difficult                                                            & 10/06/2024     & Google Scholar     \\
Towards Secure Tuning: Mitigating Security Risks Arising from Benign Instruction Fine-Tuning                         & 10/11/2024     & Google Scholar     \\
MoGU: A Framework for Enhancing Safety of Open-Sourced LLMs While Preserving Their Usability                         & 10/12/2024     & Google Scholar     \\
Buckle Up: Robustifying LLMs at Every Customization Stage via Data Curation                                          & 10/12/2024     & Google Scholar     \\
Targeted Vaccine: Safety Alignment for Large Language Models against Harmful Fine-Tuning via Layer-wise Perturbation & 10/15/2024     & Google Scholar     \\
Safety-Aware Fine-Tuning of Large Language Models                                                                    & 10/18/2024     & Google Scholar     \\
Can Editing LLMs Inject Harm?                                                                                        & 10/24/2024     & Google Scholar     \\
On the Vulnerability of Safety Alignment in Open-Access LLMs                                                         & 10/29/2024     & Google Scholar     \\
Towards Understanding the Fragility of Multilingual LLMs against Fine-Tuning Attacks                                 & 10/29/2024     & Google Scholar     \\
The effect of fine-tuning on language model toxicity                                                                 & 10/30/2024     & Google Scholar     \\
Defending Against Unforeseen Failure Modes with Latent Adversarial Training                                          & 11/12/2024     & Google Scholar     \\
The VLLM Safety Paradox: Dual Ease in Jailbreak Attack and Defense                                                   & 11/21/2024     & Google Scholar     \\
Safety Layers in Aligned Large Language Models: The Key to LLM Security                                              & 11/27/2024     & Google Scholar     \\
On Evaluating the Durability of Safeguards for Open-Weight LLMs                                                      & 11/27/2024     & Google Scholar     \\
PEFT-as-an-Attack! Jailbreaking Language Models during Federated Parameter-Efficient Fine-Tuning                     & 12/05/2024     & Google Scholar     \\
Leveraging Catastrophic Forgetting to Develop Safe Diffusion Models against Malicious Finetuning                     & 12/18/2024     & Google Scholar     \\
Separate the Wheat from the Chaff: A Post-Hoc Approach to Safety Re-Alignment for Fine-Tuned Language Models        & 12/20/2024     & Google Scholar    \\
\bottomrule
\end{tabular}
}
\end{table}

\subsection{Potential bias in qualitative analysis}
We identify certain weaknesses that might introduce bias in our qualitative analysis process (See Section \ref{survey method}). We now disclose the source of potential bias as follows:
\begin{itemize}[leftmargin=*]
\item   The paper collection process might incur potential bias.  We identify two main sources of potential bias: i) the seed papers we use in the forward snowballing might be subjective as the keywords we use for Google search might not cover all the papers.  ii) The paper collection process is executed only by the first author, which might incur human error.
\item  The initial categorization and analysis process is done by the first author and \emph{is made visible to other authors}, whose role serves as an inspector. Although the other authors double-check the correctness of the categorization, the judgment from the first author might introduce a bias to the inspectors, and this ultimately results in a bias in the categorization and analysis process.
\end{itemize}

To mitigate bias, we welcome the readers to inform authors for missing papers on relevant topics. We will update the publication list in the Github repo constantly and will also  update the pre-print version of the paper to include new uncovered research.   
\end{document}